\documentclass[11pt,a4paper]{article}
\usepackage{graphicx}
\usepackage{amsmath}
\usepackage{amsfonts}
\usepackage{natbib}
\usepackage{booktabs}
\usepackage{multirow}
\usepackage{authblk}
\usepackage[margin=.8in]{geometry}

\newcommand{\be}{\begin{equation}}
\newcommand{\ee}{\end{equation}}
\newcommand{\bea}{\begin{eqnarray}}
\newcommand{\eea}{\end{eqnarray}}
\newcommand{\mb}{\mathbf}

\newcommand{\bsb}{\boldsymbol}

\newtheorem{theo}{Theorem}

\newcommand{\betab}{\bsb\beta}
\def\qed{\mbox{\rule[0pt]{1.3ex}{1.3ex}}}

\def\SizeH{\fontsize{10pt}{10pt}\selectfont}
\def\SizeG{\fontsize{9pt}{9pt}\selectfont}
\def\SizeF{\fontsize{8pt}{8pt}\selectfont}
\def\SizeE{\fontsize{7pt}{7pt}\selectfont}

\def\SizeC{\fontsize{5pt}{5pt}\selectfont}

\input epsf

\begin{document}
\bibliographystyle{agsm}


\title{Proportional hazards model with
partly interval censoring 
and its penalized likelihood estimation
}
\date{}
\author[1]{Jun Ma \footnote{To whom correspondence should be addressed: jun.ma@mq.edu.au}}
\author[2]{Dominique-Laurent Couturier}
\author[3]{Stephane Heritier}
\author[1]{Ian Marschner}
\affil[1]{Department of Statistics, Macquarie University, Australia}
\affil[2]{School of Clinical Medicine, University of Cambridge, UK}
\affil[3]{School of Public Health and Preventive Medicine, Monash University, Australia}

\maketitle
\begin{abstract}
This paper considers the problem of semi-parametric proportional hazards model fitting for
interval, left and right
censored survival times.
We adopt a more versatile penalized likelihood method to estimate the baseline hazard and the regression
coefficients simultaneously,
where the penalty is
introduced in order to regularize the baseline hazard estimate.
We present asymptotic properties of our estimate,
allowing for the possibility that it
may lie on the boundary of the parameter space.
We also provide a computational method based on marginal likelihood, which allows
the regularization parameter to be determined automatically.
Comparisons of our method with other approaches are given
in simulations which demonstrate that our method has favourable performance. A real data application involving a model
for melanoma recurrence is
presented and an
R package implementing the methods is available.
\\[2ex]
\end{abstract}

\section{Introduction} \label{sec1:intro}
Likelihood based proportional hazard model estimation for interval censored survival data has been considered by many researchers;
see, for example, \cite{WaMcMHuQu15}, \cite{Fin86}, \cite{Sun06}, \cite{Kim03}, \cite{Pan99}, \cite{ZhHuHu10}, \cite{Huang96}, \cite{JoCoLe98} and \cite{CaiBet03}, and references therein.
In this paper, we consider the problem of fitting proportional hazards (PH) models, also known as Cox regression, where observed survival times include event times and
left, right and interval censoring times.
We present a new
MPL method which
has clear distinctions from
existing MPL methods. We develop an efficient algorithm to compute
the constrained MPL estimate and we also provide an 
accurate asymptotic covariance matrix for the MPL estimate.

Several maximum penalized likelihood (MPL) methods
have been developed to fit PH models with 
interval censoring,
such as \citet{JoCoLe98} and \citet{CaiBet03}.
These methods
estimate the baseline hazard and regression coefficients
simultaneously where the baseline hazard, 
or a function of the baseline hazard such as the
cumulative baseline hazard, is approximated by a linear combination of 
a finite number of basis functions.

Our new method is designed to address two common 
difficulties for full likelihood estimates:
(i) the baseline hazard must be
non-negatively constrained; and (ii) the asymptotic covariance matrix can be singular or near singular, leading to useless
variance estimates for the regression coefficients.
For problem (i), a common solution is to use nonnegative basis functions so that
one only constrains
the coefficients of the
basis functions
to be nonnegative.
For example, \citet{JoCoLe98} use M-spline \citep{Ramsay88} basis functions and then they express
the coefficients in the linear combination as squares.
One problem with this approach, however, is the possibility of 
instable computations.
More specifically, squaring the 
coefficients (or entire function) or expressing them using the exponential function
can turn a concave objective function into non-concave,
and therefore, create local maximums.
Problem (ii) has been addressed unsatisfactorily so far, and common approaches include
bootstrapping (e.g. \cite{Pan99}) and a method based on an efficient score function for the regression coefficients
(e.g. \cite{ChSuPe12}).
Problems with these methods are:
the bootstrapping method is time demanding,
while efficient scores are generally difficult to compute.

In this paper we first present a computationally efficient procedure for constrained MPL estimation of
the PH model, where
observations include
either fully observed event data or censored data, allowing for
left, right and interval censoring.
The nonparametric baseline hazard is approximated using
a finite number of nonnegative basis functions.
Then we develop asymptotic properties for the constrained
MPL estimates. Our asymptotic results
are novel in semi-parametric survival analysis
and they produce, according to the simulation study, accurate standard error 
estimates for both
regression coefficients and baseline hazard.

The rest of this paper is arranged as follows.
In Section \ref{sec:mpl} we formulate the problem of constrained MPL estimation for the
PH model, and then
in Section \ref{sec:algo} we present an iterative scheme for constrained MPL computations.
Asymptotic properties for the constrained MPL estimates are presented in Section \ref{sec:asymp},
with proofs given in Appendix. 
Optimal smoothing parameter selection using marginal likelihood is explained in Section \ref{sec:smth}.
Section \ref{sec:simu} reports the
results from a simulation study, and in Section \ref{sec:real} the results from a real data application are provided.
Concluding remarks are
included in Section \ref{sec:conc}.

\section{Maximum penalized likelihood formulation} \label{sec:mpl}
For individual $i$, where $i=1, \ldots, n$, let $Y_i$ be the random variable representing the time to
onset of the event of interest, and
bivariate random vector $\bsb C_i = (C_i^L, C_i^R)^T$
represents the end-points of a random censoring interval, where $C_i^L\geq 0$, $C_i^R> C_i^L$
and superscript $T$ denotes matrix transpose. Note that it is
possible for $C_i^R$ to be 
$+\infty$. We assume that $Y_i$ and $\bsb C_i$ are independent given the covariates 
and that
$Y_i$ and $\bsb C_i$ cannot be observed simultaneously. 
The observed survival time for individual $i$ is denoted by random vector
$\bsb T_i=(T_i^L, T_i^R)^T$, where 
$T_i^L=C_i^L$ and $T_i^R=C_i^R$ if $\bsb C_i$ is observed
(and thus $Y_i \in [C_i^L, C_i^R]$);
otherwise, $T_i^L=T_i^R=Y_i$ if $Y_i$ is observed.
We assume $\bsb T_i$ are independent and
values for $T_i^L$ and $T_i^R$ are denoted by $t_i^L$ and $t_i^R$ respectively.
Therefore, $(t_i^L, t_i^R, \mathbf{x}_i)$ denotes the set of available information for the $i$th individual with $i=1, \ldots, n$, and where $(t_i^L, t_i^R)$ and $\mathbf{x}_i$ respectively denote the (observed) survival time of $i$ and its covariate vector of length $p$.
If $t_i^L=0$ we have left censoring, while $t_i^R=\infty$ gives right censoring; if
$t_i^L=t_i^R$ then it represents 
an event time; for other
cases they are interval-censored observations.
For the cases of left, right
and no censoring,
since only a single time point is involved,
we can simply denote them
by a single variable $t_i$ when there is no confusion.

From the observations, we wish to estimate
the PH model
\be \label{ph}
 h(t|\mathbf{x}_i) = h_0(t)\exp(\mathbf{x}_i \bsb \beta),
\ee
where $h(t|\mathbf{x}_i)$ denotes the hazard function for
individual $i$, $h_0(t)$ represents the baseline hazard and $\bsb\beta$ is a $p$-vector of regression coefficients.
Note that $\mathbf{x}_i$ forms the $i$-th row of the
design matrix $\mathbf{X}$.
Clearly, it requires $h_0(t)\geq0$ 
so that both $h_0(t)$ and $h(t|\mathbf{x}_i)$ 
are valid hazard functions.
In order to simplify the notations below, we let
$h_i(t)=h(t|\mb{x}_i)$, $S_i(t) = S(t|\mathbf{x}_i)$ and
$H_i(t)=H(t|\mathbf{x}_i)$.

In this paper we consider the
MPL method to fit model (\ref{ph}), where $h_0(t)$ and $\bsb \beta$ are estimated simultaneously.
Since $h_0(t)$ is an infinite dimensional parameter, its estimation from
a finite number of observations is ill-conditioned.
We address this problem through approximating $h_0(t)$ using
a finite number of nonnegative basis functions,
that is
\be \label{basehazard}
 h_0(t)=\sum_{u=1}^{m} \theta_u \psi_u(t),
\ee
where $\psi_u(t)\geq 0$ are basis functions and $m$, the dimension of the approximating space, is
usually related to the number of knots defining the basis
functions. Possible choices for basis functions include indicator functions, M-splines and Gaussian density functions.
We denote the vector for distinctive knots by $\bsb\alpha$
and the number of interior distinctive knots (i.e. the distinctive knots apart from the minimum and maximum
knots) 
by $n_{\alpha}$.
The requirement that $h_0(t)\geq 0$ can now be imposed more simply through 
$\bsb \theta \geq 0$,
where $\bsb \theta$ is an $m$-vector for the $\theta_u$'s
and $\bsb \theta \geq 0$
is interpreted element-wisely. Approximation using basis functions
in PH models
has been adopted by many authors, including \cite{ZhHuHu10} for spline based sieve maximum likelihood estimation,
\cite{CaiBet03} and \cite{JoCoLe98}
for respectively penalized linear spline and M-spline based MPL estimation, and \cite{MaHeLo14} for
constrained MPL estimation.
The sieve maximum likelihood
requires the knot
sequence to grow
very slowly with
increasing sample size.
Otherwise oscillations may appear in the hazard function estimation. 
MPL is able to dampen the unpleasant oscillations.

The log-likelihood for observation $i$ is
\begin{align} \label{lliki}
 l_i(\bsb \beta, \bsb \theta) = &\ \delta_i(\log h_0(t_i)+\mathbf{x}_i\bsb \beta+\log S_i(t_i)) + \delta_i^R\log S_i(t_i) +
 \delta_i^L\log(1- S_i(t_i))\nonumber\\
  &+ \delta_i^I\log( S_i(t_i^L)-S_i(t_i^R)), 
\end{align}
where $\delta_i$ is the indicator for event times and
$\delta_i^{R}$, $\delta_i^{L}$ and $\delta_i^{I}$ are the indicators for right, left and interval censoring times respectively. 
Clearly, $\delta_i = 1-\delta_i^R-\delta_i^L-\delta_i^I$.
The log-likelihood 
from the entire data set is then
\be \label{llik}
 l(\bsb \beta, \bsb \theta) = \sum_{i=1}^n l_i(\bsb \beta, \bsb \theta).
\ee
In this paper we develop a new method to compute the penalized likelihood estimate of $\bsb \beta$ and $\bsb \theta$ 
where a penalty function is
used to smooth the $h_0(t)$ estimate.
It is a constrained optimization given by
\be \label{opt}
 (\widehat{\bsb \beta}, \widehat{\bsb \theta}) = \arg\!\max_{\hspace{-.1in}\bsb \beta, \bsb \theta}\{\Phi(\bsb \beta, \bsb \theta) = l(\bsb \beta, \bsb \theta) - \lambda J(\bsb \theta)\},
\ee
subject to $\bsb \theta \geq 0$, where $\lambda\geq 0$ is the smoothing parameter and $J(\bsb \theta)$ is a penalty function imposing smoothness on $h_0(t)$.
The well known roughness penalty (e.g. \cite{GreSil94}) takes 
$ 
 J(\bsb \theta) = \int h_0''(t)^2 dt = \bsb \theta^T \mathbf{R} \bsb \theta,
$ 
where matrix $\mathbf{R}$ has the dimension of $m \times m$ with the $(u, v)$th element
$r_{uv} = \int \psi''_u(t) \psi''_v(t) dt$.

\section{Estimation of $\bsb \beta$ and $\bsb \theta$ } \label{sec:algo}
We propose an algorithm similar to the one developed in \cite{MaHeLo14}
to find the required estimates for $\bsb \beta$ and $\bsb \theta$, where $\bsb \theta \geq 0$.
This is an alternating iterative method where each iteration involves two steps: firstly, $\bsb \beta$ is updated using the Newton algorithm, and then secondly,
$\bsb \theta$ is computed from the multiplicative iterative (MI) algorithm (e.g. 
\cite{ChMa12})
which produces estimates satisfying the nonnegative constraint.

The Karush-Kuhn-Tucker (KKT) conditions for the constrained optimization (\ref{opt}) are $\partial \Phi/\partial \beta_j = 0$, and
$\partial \Phi/\partial \theta_u=0$ if $\theta_u>0$ and $\partial \Phi/\partial \theta_u<0$ if $\theta_u=0$.
In our algorithm, 
the vector $\bsb \beta$ is first updated by the Newton algorithm at each iteration as follows:
\be \label{betaeqn}
 \bsb \beta^{(k+1)} = \bsb \beta^{(k)} + \omega_1^{(k)} \left[-
 \frac{\partial^2 \Phi(\bsb \beta^{(k)}, \bsb \theta^{(k)})}{\partial \bsb \beta \partial \bsb \beta^T}\right]^{-1}
 \frac{\partial \Phi(\bsb \beta^{(k)}, \bsb \theta^{(k)})}{\partial \bsb \beta},
\ee
where $\omega_1^{(k)}\in (0, 1]$ represents the
line search step size used to assure \newline 
$\Phi(\bsb \beta^{(k+1)}, \bsb \theta^{(k)})\geq \Phi(\bsb \beta^{(k)}, \bsb \theta^{(k)})$. 
Expression of the first and the second derivatives
are available in Appendix \ref{app0}. 
Next, $\bsb \theta$ is updated by the MI algorithm to give
\be \label{thetaeqn}
 \bsb \theta^{(k+1)} = \bsb \theta^{(k)} + \omega_2^{(k)} \mb{D}^{(k)} \frac{\partial \Phi(\bsb \beta^{(k+1)}, \bsb \theta^{(k)})}{\partial \bsb \theta},
\ee
where $\omega_2^{(k)}\in (0, 1]$ is a line search step size
and $\mb{D}^{(k)}$ is a diagonal matrix
with diagonals 
$\theta_u^{(k)}/d_u^{(k)}$ for $u = 1, \ldots, m$, where
\begin{align*}
 d_u^{(k)} = &\delta_i \Psi_u(t_i)\exp(\mathbf{x}_i\bsb \beta^{(k+1)})+\delta_i^{R} \Psi_u(t_i)
 \exp(\mathbf{x}_i\bsb \beta^{(k+1)})\\
 &+\delta_i^{I}\frac{S_i^{(k)}(t_i^L)\Psi_u(t_i^L)}{S_i^{(k)}(t_i^L)-S_i^{(k)}(t_i^R)}\exp(\mathbf{x}_i\bsb \beta^{(k+1)})
 + \lambda \left[\frac{\partial J(\bsb \theta^{(k)})}{\partial \theta_u}\right]^++\xi_u.
\end{align*}
Here, $\Psi_u(t)=\int_0^t \psi_u(w)dw$, $[a]^+ = \max\{0, a\}$ and $\xi_u$ is a nonnegative constant used to avoid the possibility of zero $d_u$; 
its choice will not affect the final solution of this algorithm.
In (\ref{thetaeqn}), the line
search step size $\omega_2^{(k)}$ is
selected such that $\Phi(\bsb \beta^{(k+1)}, \bsb \theta^{(k+1)}) \geq \Phi(\bsb \beta^{(k+1)}, \bsb \theta^{(k)})$, where equality
holds only when the algorithm has converged. Since this algorithm involves both Newton and MI
steps, we call it the Newton-MI algorithm.
Step sizes $\omega_1^{(k)}$ and $\omega_2^{(k)}$ are determined by a line search procedure.
A particular such a step size is
given by the Armijo method \citep{Armijo1966}; see also \cite{Luenberger84} for more details.

Following the same argument as in \cite{ChMa12} we can show that (i) if $\bsb \theta^{(k)}$ is nonnegative then
$\bsb \theta^{(k+1)}$ 
is also nonnegative, and (ii) under certain regularity conditions,
this Newton-MI algorithm converges to a solution satisfying the KKT conditions.

\section{Asymptotic properties} \label{sec:asymp}
\subsection{Basic formulation}
We first provide in Section \ref{sec:uncon} asymptotic
consistency for the MPL estimates of
$\bsb \beta$ and $h_0(t)$
when the number of interior distinctive knots $n_\alpha \to \infty$
but $n_\alpha/n \to 0$ and
$\lambda/n \to 0$ when $n \to \infty$. 
In Section \ref{sec:con},
asymptotic results for the constrained MPL estimates
of $\bsb\beta$ and $\bsb\theta$
are given where $\lambda = O(\sqrt{n})$. The simulation studies in Section \ref{sec:simu} reveal that
the asymptotic variances are accurate
when compared with the variances obtained from the Monte Carlo simulations.

Assuming that $h_0(t)$ is
bounded and has $r~(\geq 1)$ continuous
derivatives over $[a, b]$, and let $C^r[a, b]$ be the set for these functions, where
$\displaystyle a = \min_i t_i^L$ and
$\displaystyle b=\max_i t_i^R$.
Let the space for $\bsb\beta$ be given by
$B = \{\bsb\beta: |\beta_j|\leq C_1<\infty, \forall j\}$, a compact subset of $R^p$,
and the space for $h_0(t)$
be $A = \{h_0(t): h_0 \in C^r[a, b], 0\leq h_0(t) \leq C_2<\infty, \forall t \in [a, b]\}$.
Then the parameter space for $\bsb\tau = (\bsb\beta, h_0(t))$ is
$\Gamma = \{\bsb\tau: \bsb\beta\in B, h_0 \in A\} = B * A$.
In this section we denote the approximating function to $h_{0}(t)$ by $h_{n}(t)$:
$h_{n}(t)= \sum_{u=1}^m \theta_{un}\psi_u(t)$, where $\theta_{un}$ are assumed bounded and nonnegative
and $\psi_u(t)$ are bounded for $t\in [a, b]$; see Assumption A3 below.
Let $\bsb\theta_n = (\theta_{1n}, \ldots, \theta_{mn})^T$.
The space for
$h_n(t)$ is denoted by $A_n = \{h_n(t): 0\leq h_n(t) \leq C_3<\infty, \forall t \in [a, b]\}$.
The parameter space for $\bsb\tau_n=(\bsb\beta, h_n(t))$ is $\Gamma_n = \{\bsb \tau_n:
\bsb \beta\in B, h_n \in A_n\} = B*A_n$.
The MPL estimator of $\bsb\tau_n$
is denoted by
$\widehat{\bsb\tau}_n = (\widehat{\bsb \beta}, \widehat{h}_n(t))$, where
$\widehat{h}_n(t) = \sum_{u=1}^m \widehat{\theta}_{un}\psi_u(t)$.

Let random vectors $\bsb{W}_i =
(\delta_{i}^L, \delta_{i}^R, \delta_{i}^I, \delta_{i}, T_i^L, T_i^R, \mb{x}_i)^T$ for $i=1, \ldots, n$,
and they are assumed i.i.d. 
The density function for a $\bsb{W}_i$ is
\[
 f(\bsb{w}_i) = (h_i(t_i)S_i(t_i))^{\delta_i}(1-S_i(t_i^R))^{\delta_i^L}S_i(t_i^L)^{\delta_i^R}
 (S_i(t_i^L) - S_i(t_i^R))^{\delta_i^I}\gamma(\mb{x}_i),
\]
where $t_i^L=t_i^R=t_i$ when $\delta_i=1$ and $\gamma$ denotes the density function of $\mb{x}_i$ which
is assumed independent of $\bsb\tau$.
Let $\bsb W$ represent a general $\bsb W_i$ and
$F(\bsb w; \bsb\tau)$ be the cumulative distribution function of $\bsb W$.
Corresponding to spaces $\Gamma$ and $\Gamma_n$,
the log-likelihood function based on $\bsb W$ is denoted by $l(\bsb\tau; \bsb{W})$ and 
$l(\bsb\tau_n; \bsb{W})$ respectively.
For $\bsb\tau \in \Gamma$, define $Pl(\bsb\tau) = \int l(\bsb\tau; \bsb W) dF(\bsb W; \bsb\tau_0)=E_0(l(\bsb\tau; \bsb W))$
and $P_nl(\bsb\tau) = \frac{1}{n}\sum_{i=1}^n l(\bsb\tau; \bsb W_i)$, and for $\bsb\tau_n \in \Gamma_n$,
$Pl(\bsb\tau_n)$ and $P_nl(\bsb\tau_n)$ are similarly defined,
where $\bsb\tau_0=(\bsb \beta_0, h_{00}(t))$ denotes
the ``true'' $\bsb \tau$ which in fact maximizes $E_0(l(\bsb\tau; \bsb W))$.

Assumption A4 below assumes that for any $\bsb\tau \in B*A$,
there exist $\bsb\tau_{n}\in B*A_n$ such that 
$\rho(\bsb\tau_{n}, \bsb\tau) \to 0$ when $n \to \infty$, where
\be
 \rho(\bsb \tau_1, \bsb \tau_2) =
 \left\{\|\bsb \beta_1 - \bsb \beta_2\|_2^2 +
 \sup_{t\in[a, b]} |h_{01}(t)-h_{02}(t)|^2\right\}^{1/2}.
\ee
This assumption can be guaranteed under certain regularity conditions, such
as those in 
Proposition 2.8 in \cite{DeBoorDan74}.
Let $\mu_n=\lambda/n$. The MPL estimate $\widehat{\bsb \tau}_n$ 
maximizes $P_nl(\bsb \tau_n)-\mu_n J(\bsb \tau_n)$ for all $\bsb \tau_n \in \Gamma_n$.
According to the definition of sieve-MLE (e.g \cite{Grenander81} or \cite{WoSe91}
), our MPL estimate $\widehat{\bsb \tau}_n$, when $\mu_n \to 0$, is also
a sieve-MLE of $\bsb \tau_n$ under Assumption A4. 
This is because it satisfies
$ 
 P_nl(\widehat{\bsb \tau}_n) \geq P_nl(\bsb \tau_n^*) - \varepsilon_n,
$ 
where $\bsb\tau_n^*=\arg\max_{\bsb \tau_n \in \Gamma_n}P_nl(\bsb \tau_n)$ and
$\varepsilon_n=\mu_n|-J(\widehat{\bsb\tau}_n)+J(\bsb\tau_n^*)|
\to 0$ since
the penalty function $J$ is bounded.
Therefore, it is not a surprise that the procedures developed in \cite{WoSe91} (see also
\cite{Huang96}, \cite{ZhHuHu10} and \cite{XuLaLi04})
can be adopted to obtain asymptotic results for the MPL estimates.
We follow \cite{XuLaLi04} to develop strong consistency properties.

\subsection{
Consistency for $(\widehat{\bsb\beta}, \widehat{h}_n(t))$
when $n_{\alpha} \to \infty$ 
and $\mu_n \to 0$} \label{sec:uncon}
Here we consider the situation where $n_\alpha \to \infty$ but somewhat slower than $n \to \infty$
so that $n_{\alpha}/n \to 0$.
We further assume $\mu_n \to 0$ when $n \to \infty$.
Let $\widehat{\bsb \beta}$ and $\widehat{\bsb \theta}_n=(\widehat{\theta}_{1n}, \ldots, \widehat{\theta}_{mn})^T$ be 
the MPL estimates
of $\bsb\beta$ and $\bsb \theta_n$ 
and the corresponding baseline hazard estimate be
$\widehat{h}_n(t) = \sum_{i=1}^m \widehat{\theta}_{un} \psi_u(t)$.
We state the general consistency 
results in Theorem \ref{the2} for
estimates $\widehat{\bsb \beta}$ and $\widehat{h}_n(t)$.
These results 
require regularity conditions stated below.
\begin{enumerate}

\item[A1.]
Matrix $\mathbf{X}$ is bounded and $E(\mathbf{X}\mathbf{X}^T)$ is non-singular.

\item[A2.]
\label{A2} The penalty function $J$
is bounded over $\Gamma$ and $\Gamma_n$.

\item[A3.]
\label{A3}
For function $h_n(t)$, assume its
coefficient vector $\bsb\theta_n$ is in a compact subset of $R^m$,
and moreover,
assume its basis functions $\psi_u(t)$ are bounded for $t \in [a, b]$.

\item[A4.]
\label{A4}
The knots and
basis functions are selected in a way such that for any $h(t)\in A$ there exists a $h_{n}(t)\in A_n$ such that
$\max_t |h_{n}(t)-h(t)| \to 0$ as $n \to \infty$.

\end{enumerate}

\begin{theo}
\label{the2}
Assume Assumptions A1 -- A4
hold and
$h_0(t)$ has up to $r \geq 1$ derivatives.
Assume $n_\alpha = n^\upsilon$,
where $0<\upsilon<1$ and $\mu_n\to 0$ as $n \to \infty$.  
Then, for $n \to \infty$, \\
(1) $\|\widehat{\bsb \beta} - \bsb \beta_0\| \to 0$ almost surely, and \\
(2) $\sup_{t \in [a, b]} |\widehat{h}_n(t) - h_{00}(t)| \to 0$ almost surely. 
\end{theo}
\underline{Proof:} 
See Appendix \ref{appA}. 
\qed \\[1ex] 

If following \cite{Huang96}, \cite{ZhHuHu10} or 
\cite{XuLaLi04}, the
consistency results in Theorem \ref{the2} can be further developed to provide
rate of convergence for $\widehat{\bsb\beta}$ and $\widehat{h}_n(t)$,
and then asymptotic normality for $\widehat{\bsb\beta}$.
These results, although important theoretically,
are less useful in practice for two reasons:
(i) the covariance matrix of $\widehat{\bsb\beta}$
is difficult to compute, and (ii) the asymptotic distribution of $\widehat{\bsb\beta}$ does
not involve $\widehat{h}_n(t)$, 
making predictive inferences impractical.
We develop more useful asymptotic results below for
both $\widehat{\bsb\beta}$ and $\widehat{h}_n(t)$
assuming that
magnitude of changes of $m$
is small relative to $n$.
This assumption is equivalent to a fixed $m$. It
makes inversion of the covariance matrix of $(\widehat{\bsb\beta}, \widehat{\bsb\theta})$
feasible even when $n$ is large and therefore allows $m$ to remain in asymptotic results.
Validity of this assumption is attributed mainly to the slow convergence rate of $\widehat{h}_n(t)$. For example,
in \cite{Huang96} the rate is $n^{1/3}$ when estimating the baseline cumulative hazard $H_0(t)$
by the nonparametric estimator of \cite{GroWel92}, and in \cite{ZhHuHu10} the
rate is $n^{r/(1+2r)}$ for estimating
$\log H_0(t)$ by spline basis functions where $r$ denotes the number of bounded derivatives of
$\log H_0(t)$.

This strategy works remarkably well as demonstrated from the
simulation results in Section \ref{sec:simu}.
We furnish in the next section the
asymptotic results for constrained MPL estimates of $\bsb \beta$ and $\bsb \theta$
where $\mu_n = o(n^{-1/2})$ and $m$ is small relative to $n$.

\subsection{Asymptotic normality
when $\mu_n = o(n^{-1/2})$ and $m$ small relative to $n$
} \label{sec:con}
To simplify discussions we combine all the parameters into a single vector
$\bsb \eta = (\bsb \theta^T, \bsb \beta^T)^T$, whose length is $m+p$.
We can rewrite the penalized likelihood 
in (\ref{opt}) as
$\Phi(\bsb \eta) =
\sum_{i=1}^n\phi_i(\bsb \eta),$
where $\phi_i(\bsb \eta) = l_i(\bsb \eta) - \mu_n J(\bsb \eta)$ with $J(\bsb \eta) = J(\bsb \theta)$,
and the log-likelihood function is denoted by $l(\bsb\eta)=\sum_{i=1}^n l_i(\bsb\eta)$.
The MPL estimate of $\bsb \eta$, denoted by $\widehat{\bsb \eta}$, is
obtained by maximizing $\Phi(\bsb \eta)$
with the constraint $\bsb \theta \geq 0$.
Note that we frequently experience
active constraints (i.e. some $\theta_u=0$) when estimating $\bsb \theta$ so this fact has to be considered
when developing asymptotic results; otherwise, a non-positive definite information matrix
can be obtained.
Let $\bsb \eta_{0}$ represent the ``true value" of parameter $\bsb\eta$.
We first state 
the following assumptions needed for the asymptotics.
\begin{enumerate}
\item[B1.]
\label{B1}
Assume $\bsb{W}_i =
(\delta_{i}^R, \delta_{i}^L, \delta_{i}^I, \delta_{i}, T_i^L, T_i^R, \mb{x}_i)^T$, $i=1, \ldots, n$,
are independently and identically distributed, and the distribution of $\mb{x}_i$ is independent of $\bsb\eta$.
\item[B2.]
\label{B2}
Assume
$E_{\bsb\eta_0}[n^{-1}l(\bsb\eta)]$
exists and has a unique maximum at
$
\bsb\eta_0 \in \Omega$,
where $\Omega $ is 
the parameter set for $\bsb\tau$. Assume $\Omega$ is a compact
subspace in $R^{p+m}$.

\item[B3.]
\label{B3}
Assume $l(\bsb\eta)$ has
a finite upper bound, $l(\bsb\eta)$
is twice continuously differentiable in a neighbourhood of
$\bsb\eta_0$ and the
matrix
\be
\mathbf{G}(\bsb \eta)=\lim_{n \to \infty} n^{-1}
\frac{\partial^2 l(\bsb\eta)}{\partial \bsb\eta \partial
\bsb\eta^T} 
\ee
exists.

\item[B4.]
\label{B4}
The penalty function $J(\bsb \eta)$
is twice continuously differentiable on
$\Omega$.

\item[B5.]
\label{B5}
Let $\mb{U}$ be a matrix similar to (\ref{Umat}), which defines active constraints.
Let
$\mb{F}(\bsb\eta) = \mathbf{G}(\bsb \eta) + \mu_n \partial^2 J(\bsb\eta)/\partial \bsb\eta \partial
\bsb\eta^T.$
Assume the matrix $\mb{U}^T \mb{F}(\bsb\eta)\mb{U}$ is invertible in a neighbuorhood of $\bsb\eta_0$.
\end{enumerate}

Asymptotic properties for constrained maximum likelihood estimates
can be found in, for example, \cite{Crowder84} and \cite{MoSaKo08}, and
in the following discussions we follow more closely to the latter reference.
To elucidate discussions we assume, without loss of generality, that the
first $q$ of the $\bsb \theta\geq 0$ constraints are active in the MPL solution.
Correspondingly, define
\be \label{Umat}
 \mathbf{U} = [0_{(m-q+p)\times q}, \mathbf{I}_{(m-q+p)\times (m-q+p)}]^T,
\ee
which 
satisfies $\mathbf{U}^T\mathbf{U} = \mathbf{I}_{(m-q+p) \times (m-q+p)}$.
Now we are ready to give asymptotic results 
for the constrained MPL estimates of $\bsb \eta$.

\begin{theo}
\label{the3}
Assume Assumptions B1 -- B5 hold.
Assume there are $q$ active constraints in the MPL estimate of $\bsb \theta$
and the corresponding $\mathbf{U}$ matrix can be defined in a similar way as (\ref{Umat}).
Then, when $n \to \infty$ and $\mu_n = o(n^{-1/2})$, \\
(1) the constrained MPL estimate $\widehat{\bsb \eta}$ is consistent for $\bsb \eta_0$, and \\
(2) $\sqrt{n}(\widehat{\bsb \eta}-\bsb \eta_0)$ converges in distribution to a multivariate normal distribution  
with mean $0_{(m+p)\times 1}$ and variance matrix $\widetilde{\mathbf{F}}(\bsb \eta_0)^{-1}\mathbf{G}(\bsb \eta_0)
\widetilde{\mathbf{\mathbf{F}}}(\bsb \eta_0)^{-1}$,
where $\widetilde{\mathbf{F}}(\bsb \eta)^{-1} = \mathbf{U}(\mathbf{U}^T \mathbf{F}(\bsb \eta)
\mathbf{U})^{-1}\mathbf{U}^T$.
\end{theo}
\underline{Proof:}
See Appendix \ref{appB}. 
\qed \\[1ex] 

Note that $\mathbf{F}(\bsb \eta)-\mathbf{G}(\bsb \eta)$ converges to a zero matrix
when $n \to \infty$.
We comment that 
matrix $\widetilde{\mathbf{F}}(\bsb \eta)^{-1}$ 
is in fact very easy to compute. Firstly, $\mathbf{U}^T \mathbf{F}(\bsb \eta) \mathbf{U}$
is obtained simply by deleting the rows and columns of $\mathbf{F}(\bsb \eta)$
associated with 
the active constraints. The inverse of $\mathbf{U}^T \mathbf{F}(\bsb \eta) \mathbf{U}$ is
then calculated. Finally, $\widetilde{\mathbf{F}}(\bsb \eta)^{-1}$ is obtained by padding 
the inverse
of $\mathbf{U}^T \mathbf{F}(\bsb \eta) \mathbf{U}$ with
zeros in the 
deleted rows and columns. In practice, $\bsb\eta_0$ is unknown and the expected information
matrix $\mb{G}(\bsb\eta)$ can be difficult to compute,
we can replace $\bsb\eta_0$ by $\widehat{\bsb\eta}$ and
$\mb{G}(\bsb\eta)$ by the negative Hessian matrix.

The results in Theorem \ref{the3} are useful in practice as they accommodate 
nonzero smoothing values and
active constraints.
Moreover, inferences can be 
made with respect to, for example, regression coefficients, baseline hazard
and prediction of survival probability.
The simulation results reported in Section \ref{sec:simu} demonstrate that
biases in the MPL estimates are usually negligible when smoothing values are not large.

\section{Smoothing parameter estimation} \label{sec:smth}
Automatic smoothing parameter selection is pivotal
for successful implementation of the penalized likelihood
parameter estimation, particularly for users
who are less experienced with manual selection of smoothing values.

Existing automatic smoothing methods
fall into two main categories. Methods in the first group 
minimize model prediction
error such as Akaike's information criterion (AIC),
cross validation (CV) or generalized cross validation (GCV)
(see for example Wahba and Wold (1975) and Craven and Wahba (1979)).
Methods in the second group consider the 
penalty term as random effects (Kimeldorf and Wahba, 1970)
and treat  $\lambda$ as a variance parameter which can then be estimated by maximization of a
marginal likelihood 
or a restricted maximum likelihood (REML) (see, for instance, Wahba (1985)). For
semi-parametric PH models,
smoothing parameter selection has already been considered by, for example, 
\cite{JoCoLe98} and \cite{CaiBet03}.
We consider the marginal likelihood method in this paper.
If the marginal likelihood is difficult to obtain,
a common practice is to approximate it
using the Laplace's method (such as
Kauermann et al. (2009) and Wood (2011)).

Note that the penalty function $J(\bsb \theta) = \bsb \theta^T \mathbf{R} \bsb \theta$
can be related to a normal prior distribution for
$\bsb \theta$: 
$N(0_{m\times 1}, \sigma_{\bsb \theta}^2 \mathbf{R}^{-1})$,
where $\sigma_{\bsb \theta}^2 = 1/2\lambda$. Thus,
after omitting the terms independent of $\bsb \beta$, $\bsb \theta$ and $\sigma_{\bsb \theta}^2$,
the log-posterior 
is
\be \label{logpost}
 l_p(\bsb \beta, \bsb \theta) = -\frac{m}{2}\log \sigma_{\bsb \theta}^2 + l(\bsb \beta, \bsb \theta) - \frac{1}{2\sigma_{\bsb \theta}^2} \bsb \theta^T \mathbf{R} \bsb \theta.
\ee
The log-marginal likelihood for
$\sigma_{\bsb \theta}^2$ (after integrating out $\bsb \beta$ and $\bsb \theta$) is
\be \label{marglik}
 l_m(\sigma_{\bsb \theta}^2) = -\frac{m}{2}\log \sigma_{\bsb \theta}^2 +
 \log \int \exp\left({l(\bsb \beta, \bsb \theta)
 - \frac{1}{2\sigma_{\bsb \theta}^2} \bsb \theta^T \mathbf{R} \bsb \theta}\right) d\bsb \beta d\bsb \theta.
\ee
After applying the Laplace's approximation and plugging-in the MPL estimates for $\bsb\beta$ and $\bsb\theta$, we have
\be \label{appmarglik}
 l_m(\sigma_{\bsb \theta}^2) \approx  -\frac{m}{2}\log \sigma_{\bsb \theta}^2 + l(\widehat{\bsb \beta}, \widehat{\bsb \theta})
 -\frac{1}{2\sigma_{\bsb \theta}^2} \widehat{\bsb \theta}^T R \widehat{\bsb \theta}
 -\frac{1}{2}\log\left|\widehat{\mathbf{G}}+\mathbf{Q}(\sigma_{\bsb \theta}^2)\right|,
\ee
where $\widehat{\mathbf{G}}$ is the negative Hessian from $l(\bsb \beta, \bsb \theta)$
evaluated at $\widehat{\bsb \beta}$
and $\widehat{\bsb \theta}$ and 
\[
 \mathbf{Q}(\sigma_{\bsb \theta}^2) = \left(\begin{array}{cc}
 0 &0 \\
 0 &\frac{1}{\sigma_{\bsb \theta}^2}\mathbf{R} \end{array}\right).
\]
The solution of $\sigma_{\bsb \theta}^2$ maximizing (\ref{appmarglik}) satisfies
\be \label{smthsolu}
 \widehat{\sigma}_{\bsb \theta}^2 = \frac{\widehat{\bsb \theta}^T \mathbf{R} \widehat{\bsb \theta}}{m - \nu},
\ee
where $\nu$ is equivalent to the model degrees of freedom and is given by
$
 \nu = \text{tr}\{(\widehat{\mathbf{G}}+\mathbf{Q}(\widehat{\sigma}_{\bsb \theta}^2))^{-1} \mathbf{Q}(\widehat{\sigma}_{\bsb \theta}^2)\}.
$ 
If active constraint $\bsb \theta\geq 0$ is taken into consideration, then
$
 \nu = \text{tr}\{\mathbf{U}(\mathbf{U}^T(\widehat{\mathbf{G}}+\mathbf{Q}(\widehat{\sigma}_{\bsb \theta}^2))\mathbf{U})^{-1}\mathbf{U}^T \mathbf{Q}(\widehat{\sigma}_{\bsb \theta}^2)\}.
$ 

Since $\bsb \beta$ and $\bsb \theta$ depend on $\sigma_{\bsb \theta}^2$, the expression 
(\ref{smthsolu}) 
naturally suggests an iterative procedure:
with $\sigma_{\bsb \theta}^2$ being fixed at the its current estimate, 
the corresponding MPL estimates of $\bsb \beta$ and $\bsb \theta$ are 
obtained,
and then $\sigma_{\bsb \theta}^2$
is updated by using formula (\ref{smthsolu}) where $\widehat{\bsb \beta}$, $\widehat{\bsb \theta}$
and $\widehat{\sigma}_{\bsb \theta}^2$
on its right hand side are replaced by their most current estimates.
These iterations are continued
until the degree-of-freedom $\nu$ is stabilized.
Results in Section \ref{sec:simu} reveal that
this iterative procedure usually converges quickly
when an appropriate
knot sequence has been selected.

\section{Simulation results} \label{sec:simu}
To assess the performance of
our estimator, we performed three simulations.
The first simulation 
reproduces the simulation in \cite{CaiBet03},
which involves one covariate, a linear 
baseline hazard 
and a balanced proportions of
left, right and interval censoring. The second and third simulations consider more covariates, more complex Weibull and log-logistic baseline hazard functions, and 
unbalanced
allotments of the censored observations by censoring type.
\begin{table}[t]
\caption{Parameters used for
the simulations.
}
\label{tab-descr}
\centering
\SizeF{
\begin{tabular}{rlll}
 & \multicolumn{1}{c}{Simulation 1} & \multicolumn{1}{c}{Simulation 2}  & \multicolumn{1}{c}{Simulation 3} \\
\cmidrule(r){2-2}
\cmidrule(r){3-3}
\cmidrule(r){4-4}
  \multicolumn{4}{l}{ }  \\
\multicolumn{4}{l}{\textbf{Simulation parameters}}  \\
$\betab$ vector &  $\beta=2$  & $\betab=[0.75, -0.50,0.25]^{T}$ & $\betab=[0.25,0.25]^{T}$ \\
$\mathbf{X}$ matrix & $\mathbf{x}=[\mathbf{u}_{1}]$ & $\mathbf{X}=[\mathbf{u}_{1},5\mathbf{u}_{2},7\mathbf{u}_{3}]$ & $\mathbf{X}=[\mathbf{u}_{1},7\mathbf{u}_{2}]$ \\
$Y$ distribution & Weibull & Weibull  & Log logistic  \\
Baseline hazard & $h_{0}(y)=y$ & $h_{0}(y)=3y^{2}$  & $h_{0}(y)=\frac{4\mathrm{e}^{2}}{y(\mathrm{e}^{2}+y^{-4})}$ \\
$\gamma_{L}$ and $\gamma_{R}$  & $\gamma_{L}=1,\gamma_{R}=1$  &  $\gamma_{L}=0.9,\gamma_{R}=1.3$ & $\gamma_{L}=0.5,\gamma_{R}=1.1$ \\
 \multicolumn{4}{l}{ }  \\
\multicolumn{4}{l}{\textbf{Simulation scenarios}}  \\
Sample sizes  & $n=200,500$ & $n=100,500,2000$ & $n=100,500,2000$  \\
Percentages of events  & $\pi^{E}=0\%,10\%,25\%,50\%$ & $\pi^{E}=0\%,25\%,50\%$ & $\pi^{E}=0\%,25\%,50\%$ \\
 \multicolumn{4}{l}{ }  \\
\multicolumn{4}{l}{\textbf{Repartition of the censored observations by censoring type}}  \\
Left censoring & 32.50\% & 17.90\% & 17.90\% \\
Interval censoring & 33.00\% & 43.70\% & 60.80\% \\
Right censoring & 34.50\% & 38.40\% & 21.40\% \\
 \multicolumn{4}{l}{ }  \\
\multicolumn{4}{l}{\textbf{Specific estimator parameters}}  \\
EM I-spline & 3$^{rd}$ order I-splines& 3$^{rd}$ order I-splines& 3$^{rd}$ order I-splines \\
	& $n_\alpha=3,4$ for $n=200,500$ & $n_\alpha=3,5,7$ for $n=100,500,2000$  & $n_\alpha=3,5,7$ for $n=100,500,2000$ \\
MPL M-spline&3$^{rd}$ order M-splines& 3$^{rd}$ order M-splines& 3$^{rd}$ order M-splines \\
	& $n_\alpha=7,9$ for $n=200,500$ & $n_\alpha=7,9,11$ for $n=100,500,2000$  & $n_\alpha=7,9,11$ for $n=100,500,2000$ \\	
MPL Gaussian & $\zeta_{1}=0.35,\zeta_{2}=0.4$ & $\zeta_{1}=0.35,\zeta_{2}=0.4$  & $\zeta_{1}=0.35,\zeta_{2}=0.4$ \\
& $n_\alpha=7,9$ for $n=200,500$ & $n_\alpha=7,9,11$ for $n=100,500,2000$  & $n_\alpha=7,9,11$ for $n=100,500,2000$ \\
\hline
\end{tabular}
}
\end{table}

Observed survival time
$(T^{L}_{i}, T^{R}_{i})$, including event and censoring times, was obtained by 
\begin{align*}
T^{L}_{i}=&Y_{i}^{\delta\left(U_{i}^{E}<\pi^{E}\right)}
                  \left(\gamma_L U_{i}^{L}\right)^{\delta\left(\pi^{E}\leq U_{i}^{E},\gamma_{L}U_{i}^{L}\leq Y_{i}\leq\gamma_{R}U_{i}^{R} \right)} \left(\gamma_R U_{i}^{R}\right)^{\delta\left(\pi^{E}\leq U_{i}^{E}, \gamma_{R}U_{i}^{R}< Y_{i}\right)} \\
         &0^{\delta\left(\pi^{E}\leq U_{i}^{E},Y_{i}<\gamma_{L}U_{i}^{L}\right)},\\
T^{R}_{i}=&Y_{i}^{\delta\left(U_{i}^{E}<\pi^{E}\right)}
                   \left(\gamma_L U_{i}^{L}\right)^{\delta\left(\pi^{E}\leq U_{i}^{E},Y_{i}<\gamma_{L}U_{i}^{L}\right)}
                   \left(\gamma_R U_{i}^{R}\right)^{\delta\left(\pi^{E}\leq U_{i}^{E}, \gamma_{L}U_{i}^{L}\leq Y_{i}\leq\gamma_{R}U_{i}^{R}\right)} \\
          &\infty^{\delta\left(\pi^{E}\leq U_{i}^{E}, \gamma_{R}U_{i}^{R}< Y_{i}\right)},
\end{align*}
where $Y_{i}$ denotes the 
the event time, 
$\pi^{E}$ denotes 
the event proportion, 
${U}_{i}^{L}$, ${U}_{i}^{R}$ and $U_{i}^{E}$ denote independent standard uniform variables,
$\gamma_{L}$ and $\gamma_{R}$ are two scalars help to define interval censoring values,
and $\delta(\cdot)$ represents
the indicator function.
Note that we have adopted the convention:
$0^0=1$ and $\infty^0=1$. 

Table \ref{tab-descr} indicates the regression coefficients, $\mathbf{X}$ matrix,
baseline hazard function, 
sample size and event proportion
for each simulation. The
censoring proportion per censoring type is also indicated: Simulation 1 has
a balanced proportions,
Simulation 2 shows smaller left censoring proportions
and Simulation 3 shows larger interval censoring proportions.
The plots in
Figure \ref{fig-h0tdata} show $h_0(t)$
functions used in the
simulations.
\begin{figure}[h]
\begin{center}
\includegraphics[width=8cm]{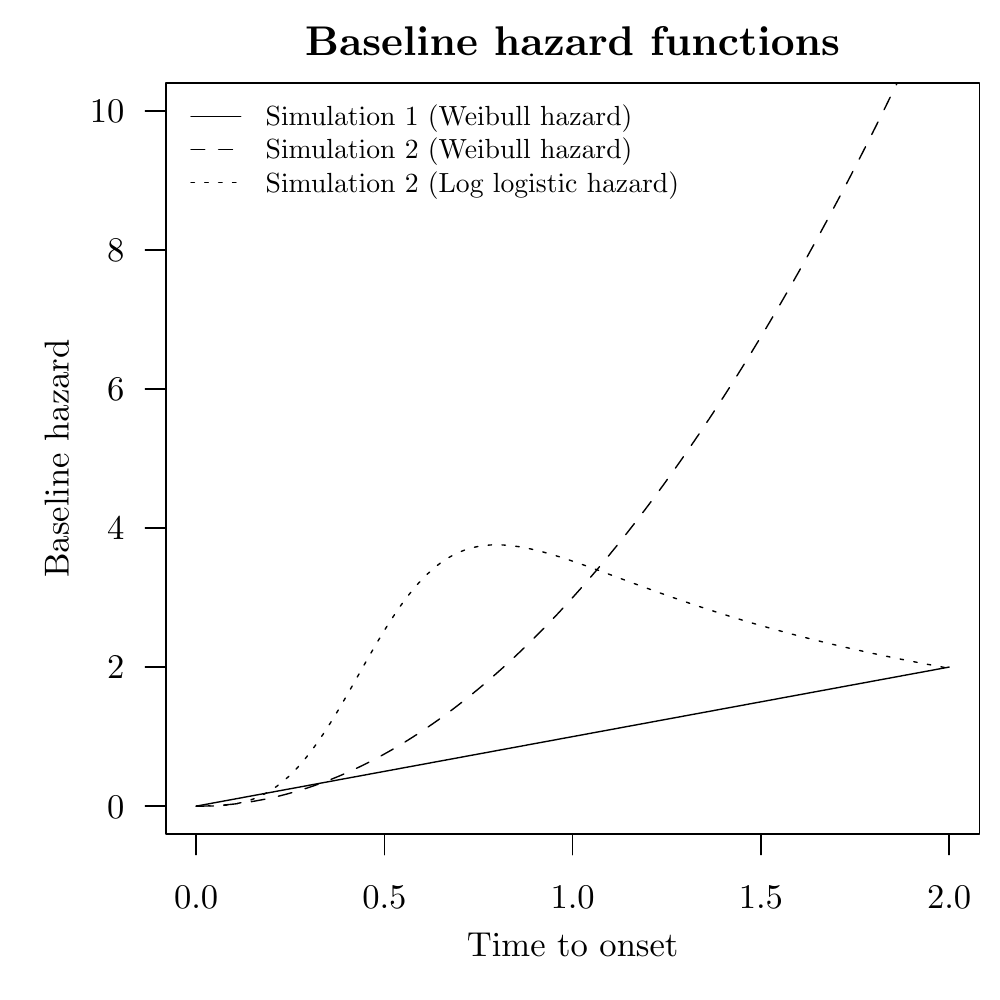} 
\caption{
Baseline hazard functions assumed in Simulations 1 to 3.
}
\label{fig-h0tdata}
\end{center}
\end{figure}

For MPL estimation, we approximate $h_{0}(t)$ by either
third order M-spline or
Gaussian basis functions,
and their expressions
are given in Appendix \ref{appC}. 
The smoothing parameter was selected automatically as described in Section \ref{sec:smth}. Our experience 
suggests that quantile based knots 
perform better than equally spaced knots, and thus this is
what we adopted in the simulations.
Let $\mathbf{a}$ be a vector for the ordered
observed survival times, including event times and
interval censoring boundaries excluding zeros and infinity values.
Then, the knots were located at equal
quantiles of $\mathbf{a}$, and the first and last knots were respectively located at the minimum and maximum of $\mathbf{a}$.
Let $n_\alpha$ be the number of interior knots.
We did not optimize $n_{\alpha}$ as from
our experience $n_{\alpha}$
has a rather low impact on the $\betab$ estimates.
Note that we used larger $n_\alpha$ for MPL than EM I-spline
as the latter suffers more numerical instability cases for larger $n_\alpha$. For
MPL, we select $n_\alpha$ roughly using cubic root of sample size $n$ but other $n_\alpha$ values
may also be used as the penalty function will restrain the MPL estimates if $n_\alpha$ is large.
The last portion of Table \ref{tab-descr} exhibits parameters used 
for M-spline and Gaussian basis. The Gaussian basis
requires an extra parameter, $\sigma^{2}_{u}$, specifying
the variance of each basis function.
We selected $\sigma^{2}_{u}$
so that the interval $[\alpha_{u}-2\sigma_{u};\alpha_{u}+2\sigma_{u}]$,
where $\alpha_u$ was a knot, would contain a given fraction
($\zeta_1$ for interior and $\zeta_2$ for boundary) knots
of $\mb{a}$. Again, MPL is less sensitive to the value of $\sigma^{2}_{u}$.
\begin{table}[h]
\caption{
Number of cases with invalid inference (due to none 
positive definite Hessian matrices) and, in parenthesis, number of cases with no solution
(due to too many planed knots), 
over 1000 estimations.}
\label{tab-num}
\centering
\SizeH{
\begin{tabular}{lccccccccc}
   & \multicolumn{2}{c}{$\pi^{E}=0\%$} &\multicolumn{2}{c}{$\pi^{E}=10\%$}  & \multicolumn{2}{c}{$\pi^{E}=25\%$} & \multicolumn{2}{c}{$\pi^{E}=50\%$} & \\
\cmidrule(r){2-3}
\cmidrule(r){4-5}
\cmidrule(r){6-7}
\cmidrule(r){8-9}
   & $n=200$ & $n=500$ & $n=200$ & $n=500$& $n=200$ & $n=500$& $n=200$ & $n=500$ & \\
   \cmidrule(r){2-2}
   \cmidrule(r){3-3}
   \cmidrule(r){4-4}
   \cmidrule(r){5-5}
   \cmidrule(r){6-6}
   \cmidrule(r){7-7}
   \cmidrule(r){8-8}
   \cmidrule(r){9-9}
\multicolumn{9}{l}{\textbf{Simulation 1}}  \\
 EM-I & 65(0) & 6(0) & 46(0) & 2(0) & 55(0) & 6(0) & 32(0) & 0(0) &  \\
 MPL-G & -(-) & -(-) & -(-) & -(-) & -(-) & -(-) & -(-) & -(-) &  \\
 \multicolumn{9}{l}{ }  \\
  \multicolumn{9}{l}{ }  \\
    & \multicolumn{3}{c}{$\pi^{E}=0\%$} & \multicolumn{3}{c}{$\pi^{E}=25\%$}  & \multicolumn{3}{c}{$\pi^{E}=50\%$}  \\
\cmidrule(r){2-4}
\cmidrule(r){5-7}
\cmidrule(r){8-10}
   & $n=100$ & $n=500$ & $n=2000$ & $n=100$ & $n=500$ & $n=2000$ & $n=100$ & $n=500$ & $n=2000$ \\
   \cmidrule(r){2-2}
   \cmidrule(r){3-3}
   \cmidrule(r){4-4}
   \cmidrule(r){5-5}
   \cmidrule(r){6-6}
   \cmidrule(r){7-7}
   \cmidrule(r){8-8}
   \cmidrule(r){9-9}
      \cmidrule(r){10-10}
\multicolumn{9}{l}{\textbf{Simulation 2}}  \\
 EM-I & 126(2) & 74(0) & 19(0) & 147(0) & 48(0) & 4(0) & 133(0) & 40(0) & 7(0) \\
 MPL-G & -(1) & -(-) & -(-) & -(-) & -(-) & -(-) & -(-) & -(-) & -(-) \\
  \multicolumn{9}{l}{ }  \\
\multicolumn{9}{l}{\textbf{Simulation 3}}  \\
 EM-I & 244(1) & 433(0) & 323(0) & 250(19) & 127(3) & 10(0) & 154(32) & 114(11) & 7(0) \\
  MPL-G & -(2) & -(-) & -(-) & -(-) & -(-) & -(-) & -(-) & -(-) & -(-) \\
   \hline
\end{tabular}
}
\end{table}

We compared the performance of 
MPL with M-spline basis (MPL-M) and
MPL with Gaussian basis (MPL-G) with some other 
semi-parametric competitors. In particular, we considered the following methods:
(i) the partial likelihood (PL) estimator with the middle point to replace left or interval censoring;
(ii) the convex minorant (CM) estimator of \cite{Pan99}, which also provides a piecewise constant estimation of the cumulative baseline hazard function;
(iii) 
the recent expectation-maximization I-spline (EM-I) estimator of \cite{WaMcMHuQu15}, which consists 
a two-stage data augmentation algorithm that exploits the relationship between the proportional hazard model and a non-homogeneous Poisson process and provides an estimation of the cumulative baseline hazard function by means of I-spline basis functions; and
(iv) 
for Simulation 1, we re-ran the simulation of \citet{CaiBet03} and therefore were able to compare
our MPL estimates with their
linear penalized spline (LPS) estimator.
For EM I-spline, we also used quantile knots
but with a reduced
number of interior knots than MPL (refer to the bottom part of Table \ref{tab-descr}) since better MSE and less numerical issues were observed.
Moreover, the Hessian matrices of EM I-spline were calculated similar to
Section \ref{sec:asymp} when zero estimates for I-spline basis parameters were obtained
as it gave better confidence interval coverages.
The partial likelihood, convex minorant and EM I-spline estimates were respectively obtained by means of the `survival', `intcox' and `ICsurv' R packages. 

We generated 1000 samples for each
combination (referred to as scenarios later) of sample size and proportion of events.
The EM I-spline estimator showed frequent
numerical issues where the causes were unclear to us.
Very rarely the MPL-G estimator also displayed
numerical issues caused by too many planned knots for the data set.
Table \ref{tab-num} reports, for each simulation, the number of cases with
non-positive definite
Hessian matrices being produced, as well as, in the parenthesis, the number of cases with no solution
due to large number of knots. For the latter case,
the number of knots was decreased until a solution could be found.
Our MPL estimator appears more reliable
as there are only 3 (over 52,000 estimations)
cases requiring adaptation of the knot sequence.
We observe that the penalty is of great help in stabilizing the MPL estimates, especially with small sample sizes.
\begin{table}[h]
\caption{Simulation 1 results for $\betab$, where $\betab=\beta_{1}=2$. Asymptotic standard errors for the \textit{convex minorant estimator} are missing due to unavailable inference for this estimator. 
}
\label{tab-sim1beta}
\centering
\SizeG{
\begin{tabular}{llrrrrrrrr}
  &  & \multicolumn{2}{c}{$\pi^{E}=0\%$} &\multicolumn{2}{c}{$\pi^{E}=10\%$}  & \multicolumn{2}{c}{$\pi^{E}=25\%$} & \multicolumn{2}{c}{$\pi^{E}=50\%$}  \\
\cmidrule(r){3-4}
\cmidrule(r){5-6}
\cmidrule(r){7-8}
\cmidrule(r){9-10}
 &   & $~n=200$ & $~n=500$ & $~n=200$ & $~n=500$& $~n=200$ & $~n=500$& $~n=200$ & $~n=500$  \\
   \cmidrule(r){3-3}
   \cmidrule(r){4-4}
   \cmidrule(r){5-5}
   \cmidrule(r){6-6}
   \cmidrule(r){7-7}
   \cmidrule(r){8-8}
   \cmidrule(r){9-9}
   \cmidrule(r){10-10}
\multicolumn{10}{l}{\textbf{Biases}}  \\
  & PL & -0.082 & -0.088 & -0.071 & -0.077 & -0.054 & -0.060 & -0.032 & -0.038 \\
   & CM & -0.031 & -0.053 &  0.162 & 0.192 &  0.143 & 0.052 & 0.008 & -0.059 \\
   & LPS &  0.023 &  0.015 &  0.034 & 0.006 & 0.020 & 0.020 & 0.020 & 0.009 \\
   & EM-I &  0.022 &  0.005 &  0.019 & 0.006 & 0.020 & 0.006 & 0.017 & 0.005 \\
   & MPL-M &  0.017 &  0.006 &  0.010 & 0.004 & 0.007 & 0.001 & 0.001 & -0.002 \\
   & MPL-G &  0.004 & -0.004 & -0.005 & -0.011 & -0.007 & -0.016 & -0.012 & -0.017 \\
   \multicolumn{10}{l}{ }  \\
  \multicolumn{10}{l}{\textbf{Mean asymptotic and (Monte Carlo) standard errors}}  \\
    &\multirow{2}{*}{PL}& 0.327\phantom{)}& 0.205\phantom{)}& 0.322\phantom{)}& 0.202\phantom{)}& 0.314\phantom{)}& 0.197\phantom{)}& 0.302\phantom{)}& 0.189\phantom{)}\tabularnewline
&&(0.320)&(0.200)&(0.315)&(0.197)&(0.307)&(0.190)&(0.302)&(0.185)\tabularnewline[1ex]
&\multirow{2}{*}{CM}&-~~~\phantom{)}&-~~~\phantom{)}&-~~~\phantom{)}&-~~~\phantom{)}&-~~~\phantom{)}&-~~~\phantom{)}&-~~~\phantom{)}&-~~~\phantom{)}\tabularnewline
&&(0.303)&(0.185)&(0.302)&(0.197)&(0.292)&(0.190)&(0.277)&(0.173)\tabularnewline[1ex]
&\multirow{2}{*}{LPS}& 0.371\phantom{)}& 0.231\phantom{)}& 0.358\phantom{)}& 0.222\phantom{)}& 0.339\phantom{)}& 0.213\phantom{)}& 0.316\phantom{)}& 0.198\phantom{)}\tabularnewline
&&(0.371)&(0.227)&(0.354)&(0.224)&(0.342)&(0.220)&(0.321)&(0.195)\tabularnewline[1ex]
&\multirow{2}{*}{EM-I}& 0.447\phantom{)}& 0.254\phantom{)}& 0.412\phantom{)}& 0.238\phantom{)}& 0.381\phantom{)}& 0.224\phantom{)}& 0.340\phantom{)}& 0.204\phantom{)}\tabularnewline
&&(0.364)&(0.228)&(0.351)&(0.219)&(0.331)&(0.208)&(0.321)&(0.196)\tabularnewline[1ex]
&\multirow{2}{*}{MPL-M}& 0.352\phantom{)}& 0.222\phantom{)}& 0.335\phantom{)}& 0.211\phantom{)}& 0.315\phantom{)}& 0.198\phantom{)}& 0.290\phantom{)}& 0.181\phantom{)}\tabularnewline
&&(0.351)&(0.220)&(0.336)&(0.210)&(0.308)&(0.194)&(0.294)&(0.180)\tabularnewline[1ex]
&\multirow{2}{*}{MPL-G}& 0.365\phantom{)}& 0.228\phantom{)}& 0.352\phantom{)}& 0.220\phantom{)}& 0.335\phantom{)}& 0.209\phantom{)}& 0.312\phantom{)}& 0.195\phantom{)}\tabularnewline
&&(0.355)&(0.224)&(0.338)&(0.214)&(0.316)&(0.202)&(0.308)&(0.191)\tabularnewline[1ex]
        \multicolumn{10}{l}{ }  \\
  \multicolumn{10}{l}{\textbf{95\% coverage probabilities}}  \\
   & PL &  0.920 &  0.852 &  0.926 &  0.867 &  0.937 &  0.901 &  0.937 &  0.933 \\
   & CM & -~~~ & -~~~ & -~~~ & -~~~ & -~~~ & -~~~ & -~~~ & -~~~ \\
   & LPS &  0.952 &  0.962 &  0.970 &  0.950 &  0.954 &  0.945 &  0.948 &  0.946 \\
   & EM-I &  0.970 &  0.967 &  0.964 &  0.961 &  0.967 &  0.966 &  0.953 &  0.950 \\
   & MPL-M &  0.951 &  0.954 &  0.957 &  0.956 &  0.960 &  0.957 &  0.949 &  0.947 \\
   & MPL-G &  0.960 &  0.958 &  0.959 &  0.953 &  0.962 &  0.959 &  0.946 &  0.947 \\
   \hline
    \end{tabular}
}
\end{table}
\begin{table}[h]
\caption{Simulation 2 results for $\betab$, where $\betab=[\beta_{1},\beta_{2},\beta_{3}]^\top=[0.75, -0.50, 0.25]^\top$. Asymptotic standard errors for the \textit{convex minorant estimator} are missing due to unavailable inference for this estimator.
}
\label{tab-sim2beta}
\centering
\SizeF{
\begin{tabular}{llrrrrrrrrr}
&    & \multicolumn{3}{c}{$\pi^{E}=0\%$} & \multicolumn{3}{c}{$\pi^{E}=25\%$}  & \multicolumn{3}{c}{$\pi^{E}=50\%$}  \\
\cmidrule(r){3-5}
\cmidrule(r){6-8}
\cmidrule(r){9-11}
&   & $n=100$ & $n=500$ & $n=2000$ & $n=100$ & $n=500$ & $n=2000$ & $n=100$ & $n=500$ & $n=2000$ \\
   \cmidrule(r){3-3}
   \cmidrule(r){4-4}
   \cmidrule(r){5-5}
   \cmidrule(r){6-6}
   \cmidrule(r){7-7}
   \cmidrule(r){8-8}
   \cmidrule(r){9-9}
   \cmidrule(r){10-10}
   \cmidrule(r){11-11}
\multicolumn{11}{l}{ }  \\
\multicolumn{11}{l}{\textbf{Biases}}  \\
  \multirow{5}{*}{$\beta_{1}$}  & PL & -0.226 & -0.240 & -0.237 & -0.160 & -0.177 & -0.174 & -0.093 & -0.115 & -0.118 \\
   & CM &  0.150 & -0.001 & -0.045 &  0.126 &  0.065 & -0.065 &  0.027 & -0.076 & -0.199 \\
   & EM-I &  0.070 &  0.006 & -0.001 &  0.050 &  0.006 &  0.003 &  0.052 &  0.007 &  0.001 \\
   & MPL-M & -0.044 & -0.029 & -0.012 & -0.078 & -0.030 & -0.010 & -0.072 & -0.027 & -0.011 \\
   & MPL-G & -0.016 & -0.029 & -0.013 & -0.033 & -0.032 & -0.013 & -0.028 & -0.027 & -0.013 \\
   \multicolumn{11}{l}{ }  \\
 \multirow{5}{*}{$\beta_{2}$}  & PL & -0.203 & -0.226 & -0.231 & -0.141 & -0.165 & -0.170 & -0.094 & -0.109 & -0.114 \\
   & CM & -0.119 & -0.372 & -0.487 & -0.377 & -0.470 & -0.481 & -0.353 & -0.400 & -0.394 \\
   & EM-I &  0.086 &  0.020 &  0.002 &  0.063 &  0.017 &  0.005 &  0.040 &  0.010 &  0.003 \\
   & MPL-M & -0.032 & -0.016 & -0.009 & -0.067 & -0.019 & -0.007 & -0.081 & -0.022 & -0.009 \\
   & MPL-G & -0.003 & -0.017 & -0.010 & -0.021 & -0.020 & -0.010 & -0.036 & -0.023 & -0.011 \\
   \multicolumn{11}{l}{ }  \\
 \multirow{5}{*}{$\beta_{3}$}  & PL & -0.207 & -0.225 & -0.235 & -0.149 & -0.168 & -0.174 & -0.090 & -0.109 & -0.117 \\
   & CM &  0.142 & -0.059 & -0.153 &  0.066 & -0.062 & -0.180 & -0.047 & -0.161 & -0.255 \\
   & EM-I &  0.081 &  0.024 & -0.001 &  0.060 &  0.016 &  0.002 &  0.050 &  0.015 &  0.002 \\
   & MPL-M & -0.027 & -0.012 & -0.011 & -0.067 & -0.020 & -0.011 & -0.074 & -0.019 & -0.011 \\
   & MPL-G &  0.002 & -0.012 & -0.013 & -0.021 & -0.021 & -0.014 & -0.029 & -0.020 & -0.013 \\
   \multicolumn{11}{l}{ }  \\
  \multicolumn{11}{l}{\textbf{Mean asymptotic and (Monte Carlo) standard errors}}  \\
  \multirow{10}{*}{$\beta_{1}$}
 &\multirow{2}{*}{PL}& 0.273\phantom{)}& 0.117\phantom{)}& 0.058\phantom{)}& 0.257\phantom{)}& 0.110\phantom{)}& 0.055\phantom{)}& 0.244\phantom{)}& 0.104\phantom{)}& 0.052\phantom{)}\tabularnewline
&&(0.286)&(0.123)&(0.060)&(0.271)&(0.115)&(0.057)&(0.257)&(0.113)&(0.054)\tabularnewline[1ex]
&\multirow{2}{*}{CM}&-~~~\phantom{)}&-~~~\phantom{)}&-~~~\phantom{)}&-~~~\phantom{)}&-~~~\phantom{)}&-~~~\phantom{)}&-~~~\phantom{)}&-~~~\phantom{)}&-~~~\phantom{)}\tabularnewline
&&(0.357)&(0.121)&(0.059)&(0.269)&(0.108)&(0.060)&(0.230)&(0.103)&(0.052)\tabularnewline[1ex]
&\multirow{2}{*}{EM-I}& 0.337\phantom{)}& 0.148\phantom{)}& 0.077\phantom{)}& 0.287\phantom{)}& 0.132\phantom{)}& 0.065\phantom{)}& 0.254\phantom{)}& 0.117\phantom{)}& 0.057\phantom{)}\tabularnewline
&&(0.369)&(0.151)&(0.073)&(0.318)&(0.131)&(0.064)&(0.282)&(0.119)&(0.056)\tabularnewline[1ex]
&\multirow{2}{*}{MPL-M}& 0.325\phantom{)}& 0.142\phantom{)}& 0.071\phantom{)}& 0.281\phantom{)}& 0.124\phantom{)}& 0.062\phantom{)}& 0.252\phantom{)}& 0.111\phantom{)}& 0.055\phantom{)}\tabularnewline
&&(0.334)&(0.145)&(0.072)&(0.280)&(0.126)&(0.063)&(0.249)&(0.114)&(0.056)\tabularnewline[1ex]
&\multirow{2}{*}{MPL-G}& 0.335\phantom{)}& 0.143\phantom{)}& 0.071\phantom{)}& 0.291\phantom{)}& 0.124\phantom{)}& 0.062\phantom{)}& 0.260\phantom{)}& 0.112\phantom{)}& 0.055\phantom{)}\tabularnewline
&&(0.333)&(0.145)&(0.072)&(0.287)&(0.126)&(0.063)&(0.257)&(0.115)&(0.056)\tabularnewline[1ex]
      \multicolumn{11}{l}{ }  \\
  \multirow{10}{*}{$\beta_{2}$}
  &\multirow{2}{*}{PL}& 0.100\phantom{)}& 0.042\phantom{)}& 0.021\phantom{)}& 0.096\phantom{)}& 0.041\phantom{)}& 0.020\phantom{)}& 0.092\phantom{)}& 0.039\phantom{)}& 0.019\phantom{)}\tabularnewline
&&(0.108)&(0.045)&(0.023)&(0.106)&(0.043)&(0.022)&(0.099)&(0.040)&(0.020)\tabularnewline[1ex]
&\multirow{2}{*}{CM}&-~~~\phantom{)}&-~~~\phantom{)}&-~~~\phantom{)}&-~~~\phantom{)}&-~~~\phantom{)}&-~~~\phantom{)}&-~~~\phantom{)}&-~~~\phantom{)}&-~~~\phantom{)}\tabularnewline
&&(0.140)&(0.066)&(0.032)&(0.104)&(0.043)&(0.021)&(0.100)&(0.040)&(0.020)\tabularnewline[1ex]
&\multirow{2}{*}{EM-I}& 0.147\phantom{)}& 0.058\phantom{)}& 0.028\phantom{)}& 0.117\phantom{)}& 0.049\phantom{)}& 0.024\phantom{)}& 0.102\phantom{)}& 0.044\phantom{)}& 0.022\phantom{)}\tabularnewline
&&(0.148)&(0.059)&(0.029)&(0.126)&(0.050)&(0.024)&(0.110)&(0.044)&(0.022)\tabularnewline[1ex]
&\multirow{2}{*}{MPL-M}& 0.118\phantom{)}& 0.053\phantom{)}& 0.027\phantom{)}& 0.102\phantom{)}& 0.046\phantom{)}& 0.024\phantom{)}& 0.091\phantom{)}& 0.042\phantom{)}& 0.021\phantom{)}\tabularnewline
&&(0.133)&(0.054)&(0.028)&(0.109)&(0.046)&(0.024)&(0.097)&(0.041)&(0.021)\tabularnewline[1ex]
&\multirow{2}{*}{MPL-G}& 0.125\phantom{)}& 0.054\phantom{)}& 0.027\phantom{)}& 0.109\phantom{)}& 0.047\phantom{)}& 0.024\phantom{)}& 0.097\phantom{)}& 0.042\phantom{)}& 0.021\phantom{)}\tabularnewline
&&(0.127)&(0.055)&(0.028)&(0.109)&(0.047)&(0.024)&(0.098)&(0.042)&(0.021)\tabularnewline[1ex]
      \multicolumn{11}{l}{ }  \\
 \multirow{10}{*}{$\beta_{3}$}
&\multirow{2}{*}{PL}& 0.069\phantom{)}& 0.029\phantom{)}& 0.015\phantom{)}& 0.065\phantom{)}& 0.028\phantom{)}& 0.014\phantom{)}& 0.062\phantom{)}& 0.026\phantom{)}& 0.013\phantom{)}\tabularnewline
&&(0.076)&(0.030)&(0.015)&(0.070)&(0.028)&(0.014)&(0.067)&(0.026)&(0.014)\tabularnewline[1ex]
&\multirow{2}{*}{CM}&-~~~\phantom{)}&-~~~\phantom{)}&-~~~\phantom{)}&-~~~\phantom{)}&-~~~\phantom{)}&-~~~\phantom{)}&-~~~\phantom{)}&-~~~\phantom{)}&-~~~\phantom{)}\tabularnewline
&&(0.082)&(0.031)&(0.017)&(0.061)&(0.026)&(0.014)&(0.059)&(0.024)&(0.013)\tabularnewline[1ex]
&\multirow{2}{*}{EM-I}& 0.082\phantom{)}& 0.042\phantom{)}& 0.021\phantom{)}& 0.075\phantom{)}& 0.038\phantom{)}& 0.017\phantom{)}& 0.065\phantom{)}& 0.033\phantom{)}& 0.015\phantom{)}\tabularnewline
&&(0.095)&(0.036)&(0.019)&(0.080)&(0.031)&(0.016)&(0.072)&(0.028)&(0.014)\tabularnewline[1ex]
&\multirow{2}{*}{MPL-M}& 0.082\phantom{)}& 0.036\phantom{)}& 0.018\phantom{)}& 0.071\phantom{)}& 0.031\phantom{)}& 0.016\phantom{)}& 0.063\phantom{)}& 0.028\phantom{)}& 0.014\phantom{)}\tabularnewline
&&(0.085)&(0.034)&(0.018)&(0.071)&(0.030)&(0.015)&(0.064)&(0.026)&(0.014)\tabularnewline[1ex]
&\multirow{2}{*}{MPL-G}& 0.086\phantom{)}& 0.036\phantom{)}& 0.018\phantom{)}& 0.074\phantom{)}& 0.032\phantom{)}& 0.016\phantom{)}& 0.066\phantom{)}& 0.028\phantom{)}& 0.014\phantom{)}\tabularnewline
&&(0.086)&(0.034)&(0.018)&(0.072)&(0.030)&(0.015)&(0.065)&(0.027)&(0.014)\tabularnewline[1ex]
      \multicolumn{11}{l}{ }  \\
  \multicolumn{11}{l}{\textbf{95\% coverage probabilities}}  \\
 \multirow{5}{*}{$\beta_{1}$}
   & PL &  0.889 &  0.647 &  0.141 &  0.907 &  0.755 &  0.344 &  0.919 &  0.840 &  0.591 \\
   & CM & -~~~ & -~~~ & -~~~ & -~~~ & -~~~ & -~~~ & -~~~ & -~~~ & -~~~ \\
   & EM-I &  0.937 &  0.939 &  0.953 &  0.931 &  0.942 &  0.954 &  0.929 &  0.934 &  0.947 \\
   & MPL-M &  0.951 &  0.945 &  0.953 &  0.955 &  0.951 &  0.945 &  0.953 &  0.938 &  0.943 \\
   & MPL-G &  0.952 &  0.943 &  0.952 &  0.959 &  0.948 &  0.946 &  0.951 &  0.941 &  0.944 \\
         \multicolumn{11}{l}{ }  \\
  \multirow{5}{*}{$\beta_{2}$}
   & PL &  0.786 &  0.242 &  0.002 &  0.841 &  0.459 &  0.018 &  0.887 &  0.692 &  0.182 \\
   & CM & -~~~ & -~~~ & -~~~ & -~~~ & -~~~ & -~~~ & -~~~ & -~~~ & -~~~ \\
   & EM-I &  0.950 &  0.949 &  0.943 &  0.933 &  0.949 &  0.946 &  0.944 &  0.950 &  0.953 \\
   & MPL-M &  0.925 &  0.945 &  0.944 &  0.927 &  0.939 &  0.934 &  0.901 &  0.937 &  0.945 \\
   & MPL-G &  0.951 &  0.946 &  0.944 &  0.950 &  0.935 &  0.935 &  0.942 &  0.938 &  0.948 \\
      \multicolumn{11}{l}{ }  \\
  \multirow{5}{*}{$\beta_{3}$}  & PL &  0.849 &  0.501 &  0.029 &  0.888 &  0.667 &  0.124 &  0.923 &  0.819 &  0.400 \\
   & CM & -~~~ & -~~~ & -~~~ & -~~~ & -~~~ & -~~~ & -~~~ & -~~~ & -~~~ \\
   & EM-I &  0.935 &  0.958 & 0.958 &  0.939 &  0.959 &  0.967 &  0.914 &  0.965 &  0.958 \\
   & MPL-M &  0.945 &  0.959 &  0.953 & 0.946 &  0.952 & 0.958 &  0.933 &  0.956 &  0.949 \\
   & MPL-G &  0.952 &  0.961 &  0.948 &  0.956 &  0.953 &  0.959 &  0.956 &  0.954 &  0.949 \\
         \hline
\end{tabular}
}

\end{table}
Tables \ref{tab-sim1beta} to \ref{tab-sim3beta}
report, for all the simulations, the biases, mean of the asymptotic
(with formula given in Theorem \ref{the3}) and Monte Carlo (displayed in brackets)
standard errors of the $\betab$ estimates.
The method of partial likelihood with mid-point imputation
displays large biases
in all the simulations, and it sometimes also produces extremely poor coverage probabilities,
such as for $\beta_2$ in Simulation 3.
Asymptotic standard error was 
not developed for the convex minorant method so that they 
are not reported here.
In Simulation 1, the MPL methods generally have the smallest biases
whilst the asymptotic standard errors of MPL and LPS
agree closely with their Monte Carlo
standard errors. Furthermore,
MPL and LPS provide the best
95\% coverage probabilities,  and we believe the less accurate
coverage probabilities of the EM I-spline method can be attributed
to its inaccurate asymptotic standard errors.
The EM I-spline estimator appears slightly too conservative when considering high percentages of censoring in Simulation 1, slightly too liberal for $\beta_{1}$ and $\beta_{3}$ when considering small sample sizes in Simulation 2, and shows poor coverage probabilities for $\beta_{2}$ for all sample sizes when considering the 100\% censoring case in Simulation 3. In general,
the coverage probabilities of MPL confidence intervals tend to
close to the 95\% nominal value in all simulations
except for $\beta_{2}$ in Simulation 2 when
sample sizes are small. 
\begin{table}[h]
\caption{Simulation 3 results for $\betab$, where $\betab=[\beta_{1},\beta_{2}]^\top=[0.25,0.25]^\top$. Asymptotic standard errors for the \textit{convex minorant estimator} are missing due to unavailable inference for this estimator.
}
\label{tab-sim3beta}
\centering
\SizeF{
\begin{tabular}{llrrrrrrrrr}
&    & \multicolumn{3}{c}{$\pi^{E}=0\%$} & \multicolumn{3}{c}{$\pi^{E}=25\%$}  & \multicolumn{3}{c}{$\pi^{E}=50\%$}  \\
\cmidrule(r){3-5}
\cmidrule(r){6-8}
\cmidrule(r){9-11}
&   & $n=100$ & $n=500$ & $n=2000$ & $n=100$ & $n=500$ & $n=2000$ & $n=100$ & $n=500$ & $n=2000$ \\
   \cmidrule(r){3-3}
   \cmidrule(r){4-4}
   \cmidrule(r){5-5}
   \cmidrule(r){6-6}
   \cmidrule(r){7-7}
   \cmidrule(r){8-8}
   \cmidrule(r){9-9}
   \cmidrule(r){10-10}
   \cmidrule(r){11-11}
\multicolumn{11}{l}{ }  \\
\multicolumn{11}{l}{\textbf{Biases}}  \\
\multirow{5}{*}{$\beta_{1}$}
  & PL & -0.426 & -0.431 & -0.439 & -0.332 & -0.357 & -0.366 & -0.275 & -0.273 & -0.282 \\
   & CM &  0.464 &  0.702 &  0.824 &  1.199 &  1.354 &  0.972 &  0.807 &  0.604 &  0.122 \\
   & EM-I & -0.110 & -0.141 & -0.186 & -0.020 &  0.008 & -0.012 &  0.010 &  0.020 &  0.001 \\
   & MPL-M & -0.016 &  0.044 &  0.018 &  0.002 &  0.045 &  0.017 & -0.010 &  0.034 &  0.011 \\
   & MPL-G & -0.028 &  0.019 &  0.005 &  0.022 &  0.025 &  0.011 & -0.003 &  0.017 &  0.004 \\
   \multicolumn{11}{l}{ }  \\
  \multirow{5}{*}{$\beta_{2}$}
   & PL & -0.444 & -0.462 & -0.463 & -0.369 & -0.387 & -0.389 & -0.263 & -0.294 & -0.299 \\
   & CM &  0.055 & -0.107 & -0.176 &  0.119 & -0.042 & -0.193 & -0.076 & -0.238 & -0.387 \\
   & EM-I & -0.098 & -0.162 & -0.201 & -0.031 & -0.021 & -0.027 &  0.024 &  0.001 & -0.008 \\
   & MPL-M & -0.038 &  0.017 &  0.006 & -0.020 &  0.013 &  0.002 & -0.004 &  0.014 &  0.001 \\
   & MPL-G & -0.043 & -0.006 & -0.005 &  0.001 & -0.005 & -0.004 & 0.012 & -0.001 & -0.004 \\
   \multicolumn{11}{l}{ }  \\
  \multicolumn{11}{l}{\textbf{Mean asymptotic and (Monte Carlo) standard errors}}  \\
    \multirow{10}{*}{$\beta_{1}$}
   &\multirow{2}{*}{PL}& 0.236\phantom{)}& 0.102\phantom{)}& 0.051\phantom{)}& 0.228\phantom{)}& 0.099\phantom{)}& 0.049\phantom{)}& 0.221\phantom{)}& 0.096\phantom{)}& 0.048\phantom{)}\tabularnewline
&&(0.257)&(0.104)&(0.053)&(0.249)&(0.101)&(0.051)&(0.234)&(0.101)&(0.050)\tabularnewline[1ex]
&\multirow{2}{*}{CM}&-~~~\phantom{)}&-~~~\phantom{)}&-~~~\phantom{)}&-~~~\phantom{)}&-~~~\phantom{)}&-~~~\phantom{)}&-~~~\phantom{)}&-~~~\phantom{)}&-~~~\phantom{)}\tabularnewline
&&(0.343)&(0.114)&(0.054)&(0.255)&(0.096)&(0.059)&(0.207)&(0.089)&(0.055)\tabularnewline[1ex]
&\multirow{2}{*}{EM-I}& 0.310\phantom{)}& 0.133\phantom{)}& 0.071\phantom{)}& 0.268\phantom{)}& 0.123\phantom{)}& 0.066\phantom{)}& 0.234\phantom{)}& 0.104\phantom{)}& 0.058\phantom{)}\tabularnewline
&&(0.311)&(0.121)&(0.058)&(0.283)&(0.123)&(0.060)&(0.256)&(0.108)&(0.054)\tabularnewline[1ex]
&\multirow{2}{*}{MPL-M}& 0.334\phantom{)}& 0.146\phantom{)}& 0.072\phantom{)}& 0.287\phantom{)}& 0.124\phantom{)}& 0.061\phantom{)}& 0.252\phantom{)}& 0.109\phantom{)}& 0.054\phantom{)}\tabularnewline
&&(0.333)&(0.147)&(0.074)&(0.284)&(0.128)&(0.062)&(0.250)&(0.110)&(0.055)\tabularnewline[1ex]
&\multirow{2}{*}{MPL-G}& 0.337\phantom{)}& 0.145\phantom{)}& 0.072\phantom{)}& 0.291\phantom{)}& 0.124\phantom{)}& 0.061\phantom{)}& 0.255\phantom{)}& 0.109\phantom{)}& 0.054\phantom{)}\tabularnewline
&&(0.325)&(0.143)&(0.073)&(0.288)&(0.125)&(0.062)&(0.251)&(0.108)&(0.054)\tabularnewline[1ex]
   \multicolumn{11}{l}{ }  \\
    \multirow{10}{*}{$\beta_{2}$}
&\multirow{2}{*}{PL}& 0.060\phantom{)}& 0.026\phantom{)}& 0.013\phantom{)}& 0.058\phantom{)}& 0.025\phantom{)}& 0.012\phantom{)}& 0.057\phantom{)}& 0.024\phantom{)}& 0.012\phantom{)}\tabularnewline
&&(0.062)&(0.026)&(0.013)&(0.063)&(0.026)&(0.013)&(0.062)&(0.026)&(0.013)\tabularnewline[1ex]
&\multirow{2}{*}{CM}&-~~~\phantom{)}&-~~~\phantom{)}&-~~~\phantom{)}&-~~~\phantom{)}&-~~~\phantom{)}&-~~~\phantom{)}&-~~~\phantom{)}&-~~~\phantom{)}&-~~~\phantom{)}\tabularnewline
&&(0.077)&(0.026)&(0.014)&(0.054)&(0.024)&(0.015)&(0.050)&(0.021)&(0.014)\tabularnewline[1ex]
&\multirow{2}{*}{EM-I}& 0.057\phantom{)}& 0.043\phantom{)}& 0.028\phantom{)}& 0.067\phantom{)}& 0.050\phantom{)}& 0.022\phantom{)}& 0.054\phantom{)}& 0.038\phantom{)}& 0.021\phantom{)}\tabularnewline
&&(0.069)&(0.032)&(0.014)&(0.072)&(0.031)&(0.016)&(0.066)&(0.028)&(0.015)\tabularnewline[1ex]
&\multirow{2}{*}{MPL-M}& 0.084\phantom{)}& 0.039\phantom{)}& 0.020\phantom{)}& 0.073\phantom{)}& 0.033\phantom{)}& 0.017\phantom{)}& 0.064\phantom{)}& 0.029\phantom{)}& 0.015\phantom{)}\tabularnewline
&&(0.088)&(0.039)&(0.020)&(0.074)&(0.033)&(0.017)&(0.064)&(0.029)&(0.015)\tabularnewline[1ex]
&\multirow{2}{*}{MPL-G}& 0.087\phantom{)}& 0.039\phantom{)}& 0.020\phantom{)}& 0.076\phantom{)}& 0.033\phantom{)}& 0.017\phantom{)}& 0.067\phantom{)}& 0.029\phantom{)}& 0.015\phantom{)}\tabularnewline
&&(0.084)&(0.038)&(0.020)&(0.076)&(0.033)&(0.017)&(0.066)&(0.028)&(0.015)\tabularnewline[1ex]
      \multicolumn{11}{l}{ }  \\
  \multicolumn{11}{l}{\textbf{95\% coverage probabilities}}  \\
    \multirow{5}{*}{$\beta_{1}$}
   & PL &  0.910 &  0.813 &  0.403 &  0.912 &  0.850 &  0.520 &  0.920 &  0.880 &  0.664 \\
   & CM & -~~~ & -~~~ & -~~~ & -~~~ & -~~~ & -~~~ & -~~~ & -~~~ & -~~~ \\
   & EM-I &  0.952 &  0.936 &  0.923 &  0.932 &  0.952 &  0.948 & 0.941 &  0.950 &  0.956 \\
   & MPL-M &  0.963 &  0.945 &  0.936 &  0.954 &  0.948 &  0.941 &  0.953 &  0.953 &  0.944 \\
   & MPL-G &  0.968 &  0.948 &  0.936 &  0.954 &  0.949 &  0.944 &  0.956 &  0.957 &  0.949 \\
         \multicolumn{11}{l}{ }  \\
    \multirow{5}{*}{$\beta_{2}$}
   & PL &  0.516 &  0.009 &  0.000 &  0.610 &  0.037 &  0.000 &  0.746 &  0.163 &  0.000 \\
   & CM & -~~~ & -~~~ & -~~~ & -~~~ & -~~~ & -~~~ & -~~~ & -~~~ & -~~~ \\
   & EM-I &  0.784 &  0.728 &  0.394 &  0.872 &  0.965 &  0.950 &  0.935 &  0.970 &  0.955 \\
   & MPL-M &  0.938 &  0.943 &  0.955 &  0.940 &  0.948 &  0.952 &  0.951 &  0.949 &  0.944 \\
   & MPL-G &  0.956 &  0.950 &  0.955 &  0.949 &  0.950 &  0.947 &  0.953 &  0.952 &  0.938 \\
       \hline
\end{tabular}
}
\end{table}

Results on $h_{0}(t)$ estimates are contained in Appendix \ref{apdx-sim}. 
For evaluating estimation of
$h_{0}(t)$, we compare MPL estimates with the
Breslow estimate for the partial likelihood method,
the piecewise constant estimate for the
convex minorant method
and the M-spline estimate for the EM I-spline method (which can be obtained by
noting the link between the M- and I-splines). 
Tables \ref{tab-sim1h0t} to \ref{tab-sim3h0t} 
report the 
biases, sample mean of the asymptotic 
standard errors (from Theorem \ref{the3}) and Monte Carlo standard errors (in bracket)
for estimating the baseline hazard function for three time values 
$t_{1}$, $t_{2}$ and $t_{3}$, respectively
corresponding to the 25th, 50th and 75th percentile of $T$.
We observe that the Breslow and convex minorant estimators provide
the largest biases and standard errors.
These tables suggest that both
MPL M-spline and MPL Gaussian estimates give reasonable
biases and small standard errors in all cases of interest.
The coverage probabilities of 95\% confidence intervals for the the baseline hazard
estimates at the chosen
percentiles of $T$
are also reported in these tables.  
No estimator performs well in all the cases, but, as expected, the coverage probabilities tend to improve when the sample sizes increase and/or the percentage of censoring decreases. In Simulation 1, the MPL M-spline and linear penalized spline estimators have
coverage probabilities close to 95\%,
while the MPL Gaussian confidence intervals are too liberal for $h_{0}(t_3)$
and the EM I-spline confidence intervals are too conservative for $h_{0}(t_1)$
and $h_{0}(t_2)$.
In Simulation 2, the coverage levels of our MPL estimators are rather poor for $h_{0}(t_3)$
especially when the sample size are small. Simulation 3 shows again poor 
coverage levels for the MPL estimators at 75th percentile 
for small sample sizes. These poor coverage probabilities are caused mainly by small standard deviations of the MPL
estimates of baseline hazards. These probabilities can be improved when using different smoothing parameters.

We also calculate and report, in these tables, 
the integrated discrepancy
between the estimated and the true
$h_{0}(t)$ over an interval $[0, t^*]$, defined as
$$
 D[\widehat{h}_{0}(t^*), h_{0}(t^*)]=\int_{0}^{t^*}\left| \widehat{h}_{0}(t)- h_{0}(t) \right| dt,
$$
where $t^{\star}$ correspond to
the 90th percentile of $T$.
Results show that both MPL estimators have much smaller integrated discrepancy than their 
competitors.

\section{
Application in a melanoma study} \label{sec:real}
In this section, we apply
the MPL estimator with M-spline bases to fit a Cox model for the time of first local melanoma recurrence for patients who were diagnosed with melanoma between 1998 and 2016 in Australia; see \cite{Mortonetal14} for some further information about a similar data set. Since our aim here is to demonstrate the MPL method in real data applications, no comparisons are made with other methods. Our data set, kindly provided by the Melanoma Institute Australia, indicates the date of melanoma diagnosis ($t_d$) and the date of last follow-up ($t_f$) with recurrence status for 2175 patients. If a melanoma recurrence was observed, it also indicates when the first recurrence was diagnosed ($t_r$) as well as the date of the last negative check before recurrence ($t_n$), if available.

Melanoma recurrence was observed for 37\% of the patients. At time of last follow-up, 70.5\% of the patients were alive and 29.5\% dead. Among the alive patients, 95\% were with no melanoma, 4\% with melanoma and 1\% with unknown melanoma status. 
Among the dead patients, 18\% were with no melanoma, 71\% with melanoma and 11\% with with unknown melanoma status. 
We set the melanoma diagnosis time as the time origin for each patient. Times of first recurrence are typically interval censored as they occurred between patient visits to the doctor. For a patient with non-missing $t_n$ and $t_r$, the first melanoma recurrence is censored in $[t_n-t_d, t_r-t_d]$. If
a patient whose $t_n$ is missing but $t_r$ is available, then
melanoma recurrence is censored in $[0, t_r-t_d]$. If $t_r$ is missing and the patient had melanoma at
$t_f$, then the recurrence time is censored in $[0, t_f-t_d]$. If $t_r$ is missing and the patient had no melanoma at $t_f$, the recurrence time is (right) censored in $[t_f-t_d,\infty)$. Cases with no observed recurrence and no known status at time of last follow up were considered as missing.
\par We considered the following covariates in our model:
(1) melanoma location at first diagnostic, a categorical variable with levels `Head and neck' (19.1\%), `Arm' (14.4\%),  `Leg' (28.7\%), `Trunk' (37.8\%);
(2) melanoma stage at first diagnostic according to Breslow's thickness scale, an ordinal variable with levels `[0,1) mm.' (15.2\%), `[1,2) mm.' (42.5\%), `[2,4) mm.' (29.2\%) and `4 mm. and more' (13.2\%);
(3) gender, a categorical variable with levels `Men' (58.1\%), `Women' (41.9\%);
(4) (centered) age in years at first diagnostic, where the range of the non centred ages is [5, 94] and the mean of non centered ages equals 55.7 years.
The contrasts were chosen so that the baseline hazard corresponds to the instantaneous risk to have a first melanoma recurrence on the head/neck for a male of 55.7 years old who was initially diagnosed with a melanoma of small size ($<$1mm). We chose to model the baseline hazard function using 
10 M-spline bases (again no effort was made to optimize this number). Two of them were placed at the extremities of the time range of interest and the others
were placed at equidistant interval mid-points. 
\begin{table}[t]
\caption{
Hazard ratio estimates ($e^{\widehat{\beta}}$), hazard ratio 95\% confidence intervals, and $p$-values of the significant tests.
}
\label{mplm_malenoma}
\centering
\begin{tabular}{llccr}
 &  & \textbf{HR estimates} & \textbf{HR 95\% CI} & \textbf{$p$-value} \\
 \cmidrule(r){3-3}
 \cmidrule(r){4-4}
 \cmidrule(r){5-5}
\textbf{Location} & Arm & 0.570 & [0.427; 0.761] & 0.0001 \\
   & Leg & 1.008 & [0.811; 1.252] & 0.9446 \\
   & Trunk & 0.802 & [0.655; 0.982] & 0.0327 \\
  \textbf{Thickness} & 1 to 2 mm. & 1.245 & [0.939; 1.650] & 0.1278 \\
   & 2 to 4 mm. & 2.390 & [1.807; 3.159] & $<$0.0001 \\
   & 4 mm. and more & 3.108 & [2.305; 4.189] & $<$0.0001 \\
  \textbf{Gender} & Female & 0.843 & [0.715; 0.993] & 0.0406 \\
  \textbf{Centered Age} (10 years) & - & 1.148 & [1.090; 1.208] & $<$0.0001 \\
     \hline
\end{tabular}

\end{table}
\begin{figure}[t]
\centering
\includegraphics[height=8cm]{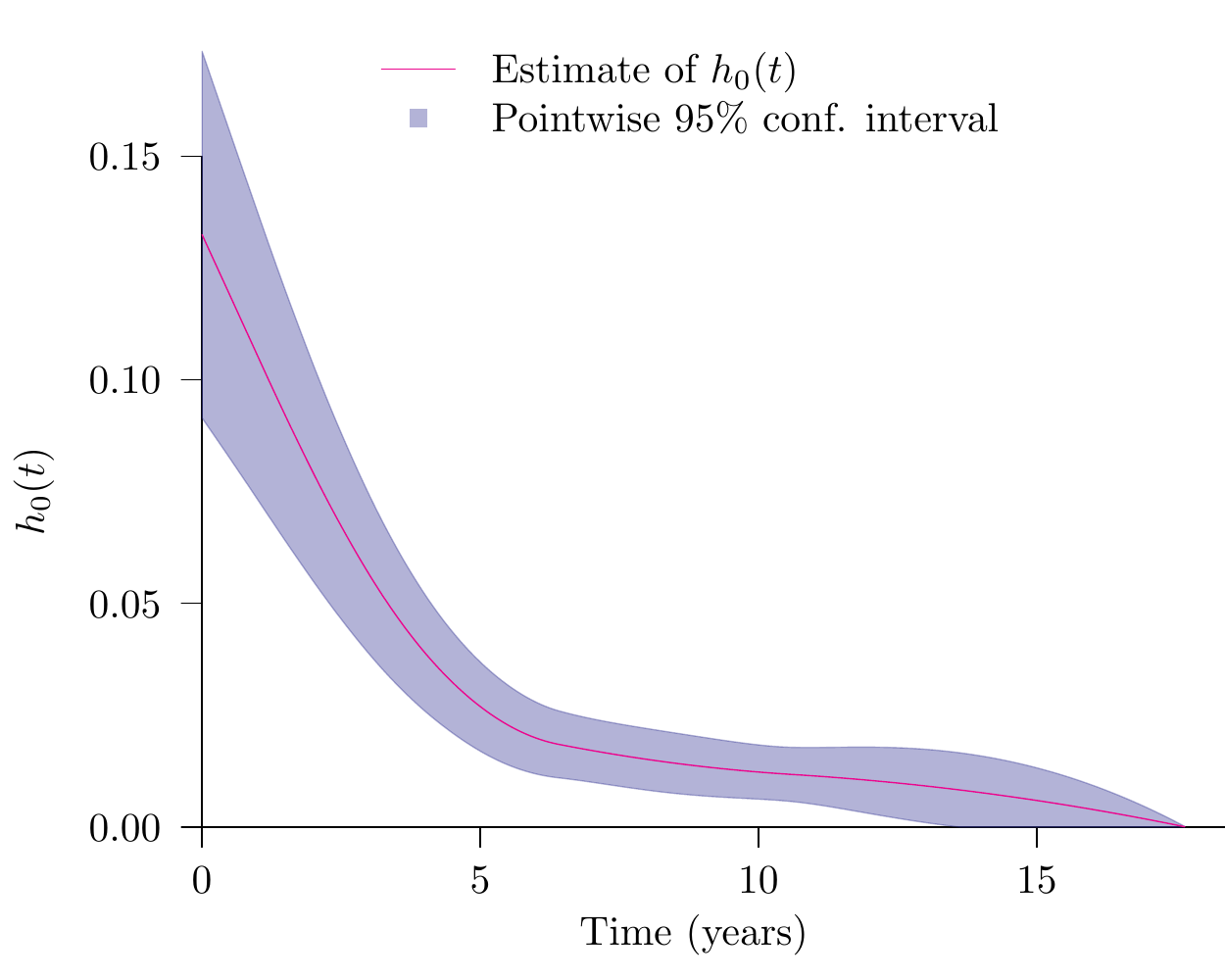}\\
\caption{Plots of baseline hazard estimate and
their 95\% CI.}
\label{melanomaplots}
\end{figure}

The hazard ratio estimates are exhibited in Table~\ref{mplm_malenoma}. Compared with melanoma that were first diagnosed at the head\&neck, melanoma at arm or trunk have significantly lower risk of recurrence. Initial melanoma thickness is another strong risk factor for melanoma recurrence.
A 10-year increase of age corresponds to a significant (9\% to 21\%) risk increase of melanoma recurrence. Gender is marginally significant, with women having a lower risk of melanoma recurrence than men.
\par The estimates of the baseline hazard function, together with its
95\% pointwise confidence interval, is displayed in Figure \ref{melanomaplots}. This plot indicates that when the covariates are all set to their baseline values, the risk of
melanoma recurrence strongly and monotonically decreases during the first 5 years. Afterwards, during the next decade, the risk continue to regularly decrease to a level close to 0.


\section{Conclusion} \label{sec:conc}
This paper develops a new approach for fitting a semi-parametric proportional hazard
model where survival time observations 
include left, interval and
right censoring times as well as event times. Since the baseline hazard is
non-parametric and subject to a nonnegativity constraint
we approximate
this function using a finite number of nonnegative basis functions and we constrain
the coefficients of the basis functions to be nonnegative. An efficient Newton-MI algorithm 
is developed.
Asymptotic results 
establish that, under certain regularity conditions,
when the number of knots 
goes to infinity (at a rate slower than 
the sample size goes to infinity) and the smoothing parameter goes to zero,
both regression coefficients and baseline hazard MPL estimates are consistent almost surely.
More practically useful asymptotic normality is 
developed and its standard errors are found to be accurate by the simulation study.
We find the MPL estimates
are capable of producing more satisfactory results than their competitors on both the regression coefficients and
baseline hazard estimates. In particular, they
produce regression coefficients with comparable (or smaller) biases and smaller 
standard errors. 
The 95\% confidence intervals from the MPL 
estimates usually achieve better coverage probabilities (i.e. closer to 95\%) than the competitors.


\section*{Appendices}
\appendix
\section{Components of score vector and Hessian matrix} \label{app0}
Let $x_{ij}$ be element $j$ of vector $\mb{x}_i$.
The first derivatives of $\Phi$ with respect to $\bsb \beta$ and $\bsb \theta$ are,
for $j = 1, \ldots, p$ and $u = 1, \ldots, m$,
\begin{align*}
\frac{\partial \Phi(\bsb \beta, \bsb \theta)}{\partial \beta_j} =&
\sum_{i=1}^n x_{ij}\left(\delta_i - \delta_i H_i(t_i) - \delta_i^{R}H_i(t_i) +
\delta_i^{L}\frac{S_i(t_i)H_i(t_i)}{1-S_i(t_i)} \right. \nonumber \\
&\left. - \delta_i^{I}\frac{S_i(t_i^L)H_i(t_i^L) - S_i(t_i^R)H_i(t_i^R)}{S_i(t_i^L)-S_i(t_i^R)}\right), \\ 
\frac{\partial \Phi(\bsb \beta, \bsb \theta)}{\partial \theta_u} =&
\sum_{i=1}^n \left(\delta_i\frac{\psi_u(t_i)}{h_0(t_i)} -\delta_i \Psi_u(t_i)e^{\mathbf{x}_i\bsb \beta}
- \delta_i^{R} \Psi_u(t_i)e^{\mathbf{x}_i\bsb \beta}+\delta_i^{L}\frac{S_i(t_i)\Psi_u(t_i)}{1-S_i(t_i)}e^{\mathbf{x}_i\bsb \beta}\right. \nonumber \\
&\left.-\delta_i^{I}\frac{S_i(t_i^L)\Psi_u(t_i^L)-S_i(t_i^R)\Psi_u(t_i^R)}{S_i(t_i^L)-S_i(t_i^R)}
 e^{\mathbf{x}_i\bsb \beta}\right)
-\lambda \frac{\partial J(\bsb \theta)}{\partial \theta_u},  
\end{align*}
where $\Psi_u(t)=\int_0^T \psi_u(\xi) d\xi$, the
cumulative of basis function $\psi_u(t)$.
Elements of the Hessian matrix are:
\begin{align*}
\frac{\partial^2 \Phi(\bsb \beta, \bsb \theta)}{\partial \beta_j \partial \beta_t} &= -\sum_{i=1}^n
 x_{ij}x_{it}\left(\delta_i H(t_i)+\delta_i^R H(t_i)+\delta_i^L\frac{S_i(t_i)H_i(t_i)(H_i(t_i)+S_i(t_i)-1)}{(1-S_i(t_i))^2}\right. \\
 &~+\left.\delta_i^I \frac{S_i(t_i^L)S_i(t_i^R)(-H_i(t_i^L)+H_i(t_i^R))^2}{(S_i(t_i^L)-S_i(t_i^R))^2}
 +\delta_i^I \frac{-S_i(t_i^R)H_i(t_i^R)+S_i(t_i^L)H_i(t_i^L)}{S_i(t_i^L)-S_i(t_i^R)}\right)\\
\frac{\partial^2 \Phi(\bsb \beta, \bsb \theta)}{\partial \beta_j \partial \theta_u} &= -\sum_{i=1}^n
x_{ij}e^{\mathbf{x}_i\betab}\left(\delta_i \Psi_u(t_i)+\delta_i^R \Psi_u(t_i)+
\delta_i^L \frac{S_i(t_i)}{1-S_i(t_i)}\left[\frac{H_i(t_i)}{1-S_i(t_i)}-1\right]\right. \\
&~+\delta_i^I \frac{S_i(t_i^L)S_i(t_i^R)(-H_i(t_i^L)+H_i(t_i^R))
(-\Psi_u(t_i^L)+\Psi_u(t_i^R))}{(S_i(t_i^L)-S_i(t_i^R))^2} \\
&~+\left.\delta_i^I \frac{S_i(t_i^L)\Psi_u(t_i^L)-S_i(t_i^R)\Psi_u(t_i^R)}{S_i(t_i^L)-S_i(t_i^R)}\right) \\
\frac{\partial^2 \Phi(\bsb \beta, \bsb \theta)}{\partial \theta_u \partial \theta_v} &= -\sum_{i=1}^n\left(
\delta_i\frac{\psi_u(t_i)\psi_v(t_i)}{h_0^2(t_i)}+e^{2\mathbf{x}_i\betab}\left[
\delta_i^L\frac{S_i(t_i)}{(1-S_i(t_i))^2}\Psi_u(t_i) \Psi_v(t_i) \right.\right. \\
&~+\left.\left. \delta_i^I \frac{S_i(t_i^L)S_i(t_i^R)}{(S_i(t_i^L)-S_i(t_i^R))^2}(\Psi_u(t_i^R)-\Psi_u(t_i^L))
(\Psi_v(t_i^R)-\Psi_v(t_i^L)) \right]\right)
\end{align*}

\section{Sketch proof of Theorem 1 
} \label{appA}
Our approach below closely follows the proofs given in \cite{XuLaLi04}, \cite{ZhHuHu10} and \cite{Huang96}.
Recall $\bsb \tau = (\bsb \beta, h_0(t)) 
\in B*A$ and $\bsb \tau_n
= (\bsb \beta, h_{n}(t)) 
\in B*A_n \subset B*A$, where spaces $B$, $A$ and $A_n$
have already been defined in Section 4. 
$h_{n}(t)$ is an approximation to $h_0(t)$ using the basis
functions. The MPL estimator is represented by $\widehat{\bsb\tau}_n$.
For $\bsb\tau_1, \bsb\tau_2 \in B*A$, define a distance measure
\[
 \rho(\bsb \tau_1, \bsb \tau_2) = \{\|\bsb\tau_1-\bsb\tau_2\|^2\}^{1/2}=\left\{\|\bsb \beta_1 - \bsb \beta_2\|_2^2 +
 \sup_{t\in[a, b]} |h_{01}(t)-h_{02}(t)|^2\right\}^{1/2},
\]
and we denote this measure by $L_2*L_{\infty}$.

The proofs below require the concept of \emph{covering number} of a space; its 
definition can be found in, for example, \cite{Pollard84}. Briefly,
this is the number of spherical balls of a given size required to cover a given space.
For a space $A$ with
measure $\kappa(A)$, we denote the covering number associated with spheral radius $\varepsilon$ by
$N(\varepsilon, A, \kappa(A))$.

Results 1 and 2 of Theorem 1 
can be demonstrated if we are able to
show that $\rho(\bsb\tau_0, \widehat{\bsb\tau}_n) \to 0$ (a.s.), where $\bsb\tau_0 = (\bsb\beta_0, h_{00}(t))$.
Since the smoothing parameter $\mu_n \to 0$ when $n \to \infty$ and the penalty function is bounded,
we can concentrated on the log-likelihood function only. The required result can be obtained through
the following results.
\begin{enumerate}
\item[(1)] Let $q(\bsb{W}; \bsb\tau)$ denote the Fr\'{e}chet 
derivative of the density functional $f(\bsb{W}; \bsb\tau)$
with respect to $\bsb\tau$. Let $\bsb\xi$ be a point in between
$\widehat{\bsb\tau}_n$ and $\bsb\tau_0$. Since $\bsb\xi$ is not the
maximum, the functional $q(\bsb{W}; \bsb\xi)$ is non-zero. Also,  
both 
$q(\bsb{W}; \bsb\xi)$ and $f(\bsb{W}; \bsb\xi)$ are bounded.
According to the definition of 
$Pl(\bsb\tau)$ 
in Section 4 and the fact $\bsb \tau_0$ maximizes $E_0 l(\bsb\tau, \bsb{W})$,  
we have
\begin{align}
 &|Pl(\widehat{\bsb\tau}_n; \bsb{W}) - Pl(\bsb\tau_0; \bsb{W})| = E_0(l(\bsb\tau_0; \bsb{W})-l(\widehat{\bsb\tau}_n; \bsb{W})) \nonumber\\
 &\geq \|f^{\frac{1}{2}}(\bsb\tau_0; \bsb{W}) - f^{\frac{1}{2}}(\widehat{\bsb\tau}_n; \bsb{W})\|_2^2
 = \left\|\frac{q(\bsb\xi; \bsb{W})}{2f^{\frac{1}{2}}(\bsb\xi; \bsb{W})}(\bsb\tau_0-\widehat{\bsb\tau}_n)\right\|_2^2 \nonumber\\
 &\geq C_4 \|\bsb\tau_0-\widehat{\bsb\tau}_n\|_2^2, \label{apeq1}
\end{align}
where the first inequality is established
since the Kullback-Leibler distance is not less than the Hellinger distance \citep{WoSh95}, the second equality
comes from the mean value theorem and $C_4$ is the lower
bound of $|q(\bsb\xi; \bsb{W})/2f^{\frac{1}{2}}(\bsb\xi; \bsb{W})|$.

\item[(2)] It then suffices to show $Pl(\widehat{\bsb\tau}_n) - Pl(\bsb\tau_0) \to 0$ almost surely. However, since 
\[
 |Pl(\widehat{\bsb\tau}_n) - Pl(\bsb\tau_0)| \leq
 |Pl(\widehat{\bsb \tau}_n)-P_nl(\widehat{\bsb \tau}_n)| + |P_nl(\widehat{\bsb \tau}_n) -Pl(\bsb\tau_0)|,
\]
we then wish to show that each term on the right hand side converges to 0 almost surely. For the first term, we just need to
implement the result from part (3) below, but the second term
demands further analyses. Define $\bsb \tau_{0n}= (\bsb\beta_0, h_{0n}(t)) \in B*A_n$, where $h_{0n}(t)$ is selected to satisfy 
$\rho(\bsb \tau_{0n}, \bsb \tau_{0}) \to 0$ (when
$n \to \infty$), which is guaranteed by Assumption 4. 
Since $\bsb \tau_0$ maximizes $Pl(\bsb\tau)$ for $\bsb \tau \in B*A$
and $\widehat{\bsb \tau}_n$ maximizes
$P_nl(\bsb \tau)$ for $\bsb \tau \in B*A_n$, we have
\[
 P_nl(\bsb \tau_{0n})- Pl(\bsb \tau_{0n}) + Pl(\bsb \tau_{0n}) - Pl(\bsb\tau_0)
 \leq P_nl(\widehat{\bsb \tau}_n) -Pl(\bsb\tau_0)\leq P_nl(\widehat{\bsb \tau}_n) -Pl(\widehat{\bsb \tau}_n).
\]
From part (3) below we have both $P_nl(\widehat{\bsb \tau}_n) -Pl(\widehat{\bsb \tau}_n)$
and $P_nl(\bsb \tau_{0n})- Pl(\bsb \tau_{0n})$ converge to 0 almost surely. $Pl(\bsb \tau_{0n}) - Pl(\bsb\tau_0)$
converges to 0 
can be established from $\rho(\bsb \tau_{0n}, \bsb \tau_{0}) \to 0$
and the fact that $l(\cdot)$ is continuous and bounded.

\item[(3)] It suffices to demonstrate $\sup_{\bsb\tau_n \in B*A_n}|P_nl(\bsb\tau_n) - Pl(\bsb\tau_n)| \to 0$ almost surely.
This can be achieved through the following steps:
    \begin{enumerate}
    \item[(i)] Firstly, we show that $N(\varepsilon, A_n, L_{\infty}) \leq (6C_5C_6/\varepsilon)^m$ where
    constants $C_5$ and $C_6$ will be specified below. This is because for any
    $h_{n1}, h_{n2} \in A_n$ (hence each $h_{nj}(t) = \sum_u \theta_{un}^j\psi_u(t)$),
    $\max_t |h_{n1}(t)-h_{n2}(t)| \leq C_5 \max_u|\theta_{un}^1-\theta_{un}^2| \leq C_5 \|\bsb\theta^1 - \bsb\theta^2 \|_2 
    $, where $C_5$ is 
    the upper bound of $\sum_u\psi_u(t)$ and $\bsb\theta^j$ is an $m$-vector with elements $\theta_{un}^j$.
    Thus, 
    $N(\varepsilon, A_n, L_{\infty}) \leq N(\varepsilon/C_5, \{0\leq \theta_{un}\leq C_6,
    1\leq u\leq m\}, L_2) \leq (6C_5C_6/\varepsilon)^m$, and the last inequality comes from Lemma 4.1 of \cite{Pollard84}.
    \item[(ii)] Secondly we wish to demonstrate that $N(\varepsilon, {\cal L}_n, L_{\infty}) \leq K/\varepsilon^{p+m}$, where the constant
    $K$ will be explicated below and ${\cal L}_n = \{l(\bsb\tau_n), \bsb\tau_n \in B*A_n\}$. In fact, $N(\varepsilon, {\cal L}_n, L_{\infty}) \leq N(\varepsilon/2, B, L_2)\cdot
    N(\varepsilon/2, A_n, L_{\infty}) \leq (12C_1/\varepsilon)^p(12C_5C_6/\varepsilon)^m = K/\varepsilon^{p+m}$, where $C_1$ is the upper bound of $\{|\beta_j|, \forall j\}$ and
    $K = 12^{p+m}C_1^p(C_5C_6)^m$.
    \item[(iii)] Select $\alpha_n = n^{-1/2+\phi_1} \sqrt{\log n}$ where $\phi_1 \in (\phi_0/2, 1/2)$ with $\phi_0 < 1$, and
    define $\varepsilon_n = \varepsilon \alpha_n$. Following
    Theorem 1 of \cite{XuLaLi04} we can show that
    $\text{var}[P_n l(\bsb\tau_n)]/(8 \varepsilon_n^2) \to 0$ when $n \to \infty$
    for any $\bsb \tau_n \in B*A_n$.
    \item[(iv)] Finally, from the result II.31 of \cite{Pollard84}, we have 
    \begin{align*}
     &P[\sum_{B*A_n}|P_n l(\bsb\tau_n)-P_n l(\bsb\tau)|>8\varepsilon_n]\leq 8N(\varepsilon_n, {\cal L}_n, L_{\infty})e^{-\frac{1}{128}n\varepsilon_n^2}\\
     &\leq 8K e^{-\left(\frac{1}{128}\varepsilon^2 + \frac{p+m}{2}\frac{\log(\varepsilon^2n^{-1+2\phi_1}\log n)}{n^{2\phi_1}\log n}\right)n^{2\phi_1}\log n},
    \end{align*}
    which converges to zero as the second term in bracket of the exponential function goes to zero when $n\to \infty$.
    Therefore,
    by the Borel-Cantelli lemma, we have $\sup_{B*A_n} |P_nl(\bsb\tau_n) - Pl(\bsb\tau_n)| \to 0$ almost surely.

    \end{enumerate}
%

\end{enumerate}

\section{Sketch proof of Theorem 2 
} \label{appB}
Let $\overline{l}(\boldsymbol\eta)=E_{\boldsymbol\eta_0}[n^{-1}l(\boldsymbol\eta)]$.
It follows from the 
strong law of large numbers that $n^{-1}l(\boldsymbol\eta) \rightarrow
\overline{l}(\boldsymbol\eta)$ almost surely and uniformly 
for $\boldsymbol\eta \in \Omega$.
This result, together with $\mu_n \to 0$ as $n \to \infty$ and
$\boldsymbol\eta_0$ being the unique maximum of
$\overline{l}(\boldsymbol \eta)$ due to Assumption B2, implies that $\widehat{\boldsymbol\eta} \rightarrow \boldsymbol\eta_0$ almost surely by applying, for example, Corollary 1 of \cite{HoPo94}.

Next we 
prove the
asymptotic normality result.
From the KKT necessary conditions (6) and (7) 
we have that
the constrained MPL estimate $\widehat{\bsb\eta}$ satisfies 
\be
 \mb{U}^T \frac{\partial \Phi(\widehat{\bsb\eta})}{\partial\bsb\eta}  = 0,
\ee
where $\mb{U}$ is defined in equation (10) of section 4.3.
According to the Taylor expansion
\begin{equation} \label{tye}
  \frac {\partial \Phi(\widehat{\boldsymbol\eta})}{\partial\boldsymbol\eta}=
    \frac {\partial \Phi(\boldsymbol\eta_0)} {\partial\boldsymbol\eta} +
    \frac {\partial^2 \Phi(\widetilde{\boldsymbol\eta})} {\partial\boldsymbol\eta\partial\boldsymbol\eta^{\text{T}}}
    (\widehat{\boldsymbol\eta} - \boldsymbol \eta_0),
\end{equation}
where $\widetilde{\boldsymbol\eta}$ is a vector between $\widehat{\boldsymbol\eta}$ and $\boldsymbol\eta_0$.
Therefore
\be \label{ncond}
 0 = \mb{U}^T\frac {\partial \Phi(\bsb\eta_0)} {\partial\bsb\eta} +
    \mb{U}^T\frac {\partial^2 \Phi(\widetilde{\bsb\eta})} {\partial\bsb\eta\partial\bsb\eta^{\text{T}}}
    (\widehat{\bsb\eta} - \bsb\eta_0).
\ee
Next, let 
$\widehat{\bsb\chi} $ be $\widehat{\bsb\eta}$ after deleting the active constraints and 
$\bsb\chi_0$ be similarly defined corresponding to $\bsb\eta_0$, then
\be \label{Utye}
 \widehat{\bsb\eta} - \bsb\eta_0 = \mb{U}(
 \widehat{\bsb\chi}-\bsb\chi_0).
\ee
Substituting (\ref{Utye}) into (\ref{ncond}), solving for $ \widehat{\bsb\chi}-\bsb\chi_0$
and then using (\ref{Utye}) to convert the result back to 
$\widehat{\bsb\eta} - \bsb\eta_0$ again, we eventually have
\be \label{CLTeta}
 \sqrt{n} (\widehat{\bsb\eta} - \bsb\eta_0)=-\mb{U}\left(\mb{U}^T\frac{1}{n}\frac {
   \partial^2 \Phi(\widetilde{\bsb\eta})}
   {\partial\bsb\eta\partial\bsb\eta^T} \mb{U}\right) ^{-1} \mb{U}^T \left(\frac{1}{\sqrt{n}}
  \frac {\partial l (\bsb\eta_0)} {\partial \bsb\eta}+o(1)\right).
\ee
In (\ref{CLTeta}), $-n^{-1}\partial^2 \Phi(\bsb\eta)/\partial\bsb\eta\partial\bsb\eta^T$ converges
to $F(\bsb\eta)$ (a.s.) by the law of large numbers. Also note that $\widetilde{\bsb\eta} \to \bsb\eta_0$ almost surely.
Then after applying the central limit theorem to $n^{-1/2}\partial l (\bsb\eta_0)/\partial \bsb\eta$
we have the required asymptotic normality result.

\section{M-spline and Gaussian basis functions}\label{appC}
If the baseline hazard is approximated by 
Gaussian basis functions, $\psi_{u}(t)$ is a truncated Gaussian distribution with location parameter $\alpha_{u}$ (which are knots), scale parameter $\sigma_{u}$ and range $[t_{(1)},t_{(n)}]$. This leads to the following expressions of $\psi_{u}(t)$ and
its cumulative function $\Psi_{u}(t)$:
\begin{align*}
\psi_{u}(t) &= \frac{1}{\sigma_{u}\delta_{u}} \phi\left(\frac{t-\alpha_{u}}{\sigma_{u}}\right),~
\\
\Psi_{u}(t) &= \int\limits_{t_{(1)}}^{t}\psi_{u}(v)dv = \frac{1}{\Delta_{u}}\left[ \Phi\left(\frac{t-\alpha_{u}}{\sigma_{u}}\right)- \Phi\left(\frac{t_{(1)}-\alpha_{u}}{\sigma_{u}}\right)\right],
\end{align*}
where
$t_{(1)}\leq t \leq t_{(n)}$,
$\phi(\cdot)$ and $\Phi(\cdot)$ respectively are the density and cumulative density functions of the standard Gaussian distribution,
$\Delta_{u}=\Phi((t_{(n)}-\alpha_{u})/\sigma_{u}) - \Phi((t_{(1)}-\alpha_{u})/\sigma_{u})$
and
$\alpha_{u}\in \mathbb{R}$ and $\sigma_{u}>0$. 
%
If the baseline hazard is approximated by means of M$-$spline basis functions of order $o$ \citep{Ramsay88}, we get the following expressions of $\psi^{o}_{u}(t)$ and $\Psi^{o}_{u}(t)$:
\begin{align*}
\psi^{o}_{u}(t) &=
\begin{cases}
\frac{\delta(\alpha^{\star}_{u}\leq t < \alpha^{\star}_{u+1})}{\alpha^{\star}_{u+1}-\alpha^{\star}_{u}} & \text{if } o=1, \\
\frac{o}{o-1} \frac{\delta(\alpha^{\star}_{u}\leq t < \alpha^{\star}_{u+o})}{\alpha^{\star}_{u+o} - \alpha^{\star}_{u}}
		\left[\left(t-\alpha^{\star}_{u} \right)\psi^{o-1}_{u}(t) + \left(\alpha^{\star}_{u+o}-t \right)\psi^{o-1}_{u+1}(t) \right] & \text{otherwise},
\end{cases}\label{eq:psi:mspline} \\
\Psi^{o}_{u}(t) &= 
\delta(\alpha^{\star}_{u}>t)\left[\sum\limits_{v=u+1}^{\text{min}(u+o,m+1)}
\frac{\alpha^{\star\star}_{v+o+1}-\alpha^{\star\star}_{v}}{o+1}\psi^{o+1}_{v}(t)\right]^{\delta(\alpha^{\star}_{u}<t<\alpha^{\star}_{u+o})},
\end{align*}
where
$t_{(1)}\leq t \leq t_{(n)}$,
$\alpha_{u}$ is the $u$th element of knots vector $\bsb\alpha$ 
whose length is $n_{\alpha}$ ($\alpha_{u} \in \mathbb{R}$ and $\alpha_{u}<\alpha_{u+1}$),
$m = n_{\alpha}+o-2$,
$\bsb\alpha^{\star}=\left[\min(\bsb\alpha)\mathbf{1}^T_{o-1},\bsb\alpha^T, \max(\bsb\alpha)\mathbf{1}^T_{o-1}\right]^T$,
and $\bsb\alpha^{\star\star}=\left[\min(\bsb\alpha)\mathbf{1}^T_{o},\bsb\alpha^T, \max(\bsb\alpha)\mathbf{1}^T_{o}\right]^T$,
where $\mathbf{1}_{o}$ denotes
a vector of 1 of length $o$.
Note that $\Psi^{o}_{u}(t)$, the cumulative function 
of $\psi^{o}_{u}(t)$, is referred to as an I$-$spline. M$-$spline basis functions have the following 
properties: $\int_{-\infty}^{\alpha^{\star}_{u}}\psi^{o}_{u}(v)dv=0$, $\int_{\alpha^{\star}_{u}}^{\alpha^{\star}_{u+o}}\psi^{o}_{u}(v)dv=1$ and $\int_{\alpha^{\star}_{u+o}}^{\alpha^{\star}_{\infty}}\psi^{o}_{u}(v)dv=0$.

\section{Simulation results related to baseline hazard functions}\label{apdx-sim}
In this section we present simulation results regarding to $h_0(t)$ from the three simulations. Specifically,
tables \ref{tab-sim1h0t} -- \ref{tab-sim3h0t} report biases, mean asymptotic and Monte Carlo
standard errors and 95\% coverage probabilities, all at 25\%, 50\% and 75\% quantile points of $h_0(t)$. Figure 1
displays the true $h_0(t)$, the average MPL $h_0(t)$ estimates and pointwise 95\% confidence intervals.

\begin{figure} [h]
\begin{center}
\includegraphics[width=15cm]{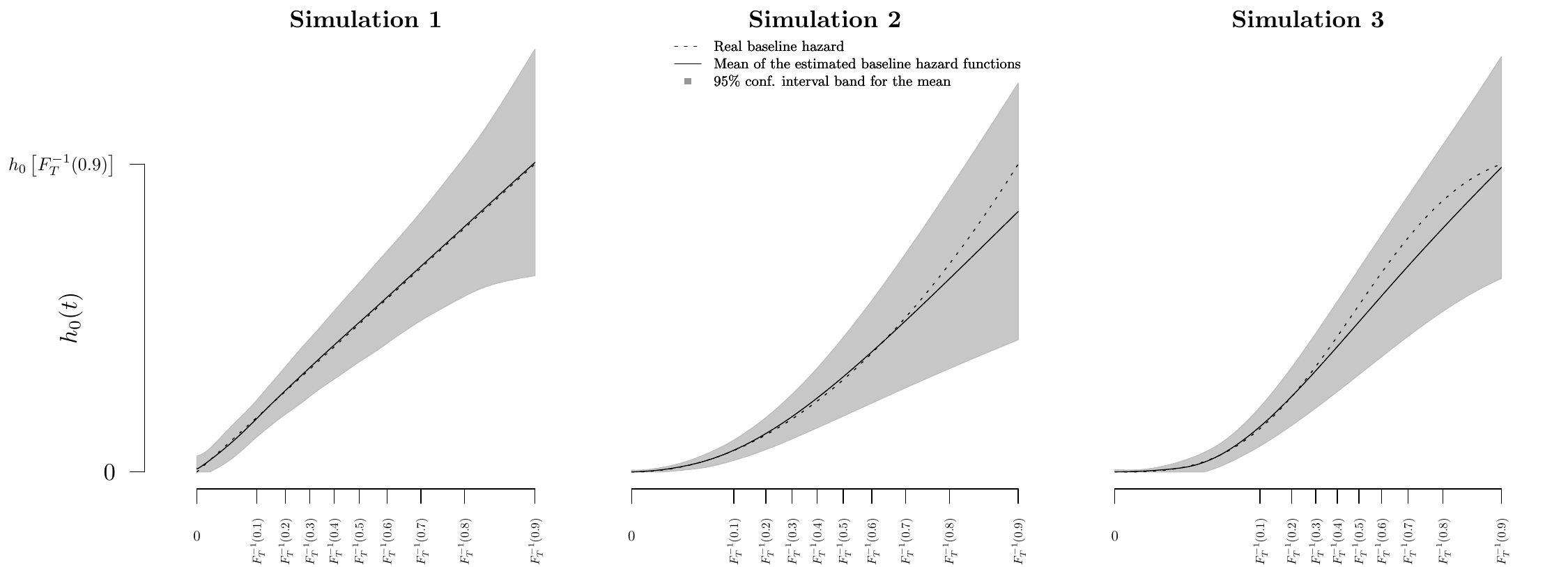} 
\caption{
Mean baseline hazard estimates and their pointwise 95\% confidence intervals.
}
\label{fig-h0tdata}
\end{center}
\end{figure}

\begin{table} 
\centering
\SizeE{
\begin{tabular}{llrrrrrrrr}
  &  & \multicolumn{2}{c}{$\pi^{E}=0\%$} &\multicolumn{2}{c}{$\pi^{E}=10\%$}  & \multicolumn{2}{c}{$\pi^{E}=25\%$} & \multicolumn{2}{c}{$\pi^{E}=50\%$}  \\
 &   & $n=200$ & $n=500$ & $n=200$ & $n=500$& $n=200$ & $n=500$& $n=200$ & $n=500$  \\
\multicolumn{10}{l}{Biases}  \\
 \SizeC{\multirow{6}{*}{$h_{0}(t_{1})$}}
 & Breslow &  4.982 &  4.918 &  4.608 &  4.517 &  4.258 &  4.294 &  3.877 &  3.661 \\
  & CM &  0.182 &  0.197 & -0.300 & -0.449 & -0.430 & -0.436 & -0.296 & -0.304 \\
 & LPS & -0.096 & -0.062 & -0.101 & -0.048 & -0.073 & -0.042 & -0.053 & -0.027 \\
  & EM-I & -0.032 & -0.100 & -0.027 & -0.064 & -0.021 & -0.021 & -0.016 & -0.002 \\
 & MPL-M &  0.014 &  0.014 &  0.019 &  0.009 &  0.021 &  0.009 &  0.026 &  0.011 \\
  & MPL-G &  0.007 &  0.005 &  0.017 &  0.016 &  0.022 &  0.027 &  0.025 &  0.027 \\
  \multicolumn{10}{l}{ }  \\
 \SizeC{\multirow{6}{*}{$h_{0}(t_{2})$}}
 & Breslow &  1.180 &  1.113 &  1.403 &  1.250 &  1.535 &  1.586 &  1.490 &  1.798 \\
 & CM &  0.189 &  0.064 & -0.162 & -0.192 & -0.160 & -0.197 & -0.147 & -0.194 \\
 & LPS &  0.006 &  0.014 & -0.013 &  0.020 &  0.013 &  0.006 &  0.018 &  0.012 \\
 & EM-I &  0.011 &  0.131 &  0.021 &  0.081 &  0.018 &  0.027 &  0.020 &  0.008 \\
 & MPL-M &  0.028 &  0.012 &  0.029 &  0.010 &  0.026 &  0.009 &  0.025 &  0.011 \\
 & MPL-G & -0.009 &  0.015 &  0.006 &  0.027 &  0.006 &  0.023 &  0.011 &  0.024 \\
  \multicolumn{10}{l}{ }  \\
 \SizeC{\multirow{6}{*}{$h_{0}(t_{3})$}}
 & Breslow &  4.256 &  2.740 &  4.037 &  2.627 &  2.394 &  2.220 &  1.900 &  2.186 \\
 & CM &  0.085 & -0.092 &  0.037 & -0.084 &  0.020 & -0.142 & -0.045 & -0.097 \\
   & LPS &  0.052 &  0.000 &  0.019 &  0.005 &  0.018 & -0.001 &  0.005 &  0.000 \\
   & EM-I & -0.016 & -0.064 & -0.038 & -0.051 & -0.039 & -0.022 & -0.028 & -0.003 \\
 & MPL-M &  0.020 &  0.007 &  0.025 &  0.014 &  0.025 &  0.013 &  0.025 &  0.013 \\
  & MPL-G & -0.080 & -0.037 & -0.065 & -0.017 & -0.043 &  0.007 & -0.012 &  0.018 \\
    \multicolumn{10}{l}{ }  \\
   \multicolumn{10}{l}{Mean asymptotic (Monte Carlo) standard errors}  \\
  \SizeC{\multirow{12}{*}{$h_{0}(t_{1})$}}
   & Breslow & -~~~ & -~~~ & -~~~ & -~~~ & -~~~ & -~~~ & -~~~ & -~~~\\
   &    & (4.521) &  (3.583) &  (3.889) &  (3.471) &  (4.539) &  (3.622) &  (4.724) &  (3.556) \\[1ex]
   & CM & -~~~ & -~~~ & -~~~ & -~~~ & -~~~ & -~~~ & -~~~ & -~~~\\
   &   &  (0.666) &  (0.812) &  (0.367) &  (0.430) &  (0.321) &  (0.286) &  (0.317) &  (0.359) \\[1ex]
   & LPS &  0.106 &  0.075 &  0.101 &  0.071  &  0.098 &  0.067 &  0.094 &  0.064 \\
   &   &  (0.112) &  (0.081) &  (0.108) &  (0.076) &  (0.101) &  (0.071) &  (0.100) &  (0.064) \\[1ex]
   & EM-I &  0.150 &  0.136 &  0.130 &  0.102 &  0.121 &  0.089 &  0.108 &  0.078 \\
   &   &  (0.122) &  (0.082) &  (0.119) &  (0.081) &  (0.111) &  (0.078) &  (0.104) &  (0.070) \\[1ex]
   & MPL-M &  0.093 &  0.061 &  0.089 &  0.056 &  0.083 &  0.0512&  0.076 &  0.047 \\
   &   &  (0.101) &  (0.064) &  (0.094) &  (0.058) &  (0.084) &  (0.050) &  (0.077) &  (0.046) \\[1ex]
   & MPL-G &  0.113 &  0.073 &  0.111 &  0.072 &  0.107 &  0.070 &  0.101 &  0.066 \\
   &   &  (0.115) &  (0.071) &  (0.113) &  (0.073) &  (0.107) &  (0.068) &  (0.102) &  (0.064) \\[1ex]
    \multicolumn{10}{l}{ }  \\
  \SizeC{\multirow{12}{*}{$h_{0}(t_{2})$}}
   & Breslow & -~~~ & -~~~ & -~~~ & -~~~ & -~~~ & -~~~ & -~~~ & -~~~\\
   &   &  (3.717) &  (2.525) &  (4.719) &  (2.667) &  (3.538) &  (3.027) &  (2.856) &  (6.237) \\[1ex]
   & CM & -~~~ & -~~~ & -~~~ & -~~~ & -~~~ & -~~~ & -~~~ & -~~~ \\
   &   &  (1.384) &  (1.379) &  (0.790) &  (0.776) &  (0.667) &  (0.590) &  (0.741) &  (0.552) \\[1ex]
   & LPS &  0.173 &  0.120 &  0.163 &  0.113 &  0.157 &  0.106 &  0.149 &  0.099 \\
   &   &  (0.171) &  (0.120) &  (0.170) &  (0.109) &  (0.160) &  (0.101) &  (0.152) &  (0.092) \\[1ex]
   & EM-I &  0.265 &  0.265 &  0.218 &  0.189 &  0.195 &  0.150 &  0.167 &  0.122 \\
   &   &  (0.183) &  (0.168) &  (0.176) &  (0.152) &  (0.165) &  (0.133) &  (0.153) &  (0.118) \\[1ex]
   & MPL-M &  0.149 &  0.095 &  0.141 &  0.088 &  0.131 &  0.082 &  0.119 &  0.074 \\
   &   &  (0.152) &  (0.095) &  (0.148) &  (0.088) &  (0.131) &  (0.079) &  (0.119) &  (0.073) \\[1ex]
   & MPL-G &  0.157 &  0.103 &  0.156 &  0.103 &  0.151 &  0.100 &  0.143 &  0.094 \\
   &   &  (0.165) &  (0.103) &  (0.163) &  (0.104) &  (0.151) &  (0.096) &  (0.142) &  (0.093) \\[1ex]
    \multicolumn{10}{l}{ }  \\
 \SizeC{\multirow{12}{*}{$h_{0}(t_{3})$}}
  & Breslow & -~~~ & -~~~ & -~~~ & -~~~ & -~~~ & -~~~ & -~~~ & -~~~ \\
   &   &  (68.918) & (22.245) & (58.760) & (19.951) &  (6.620) &  (6.557) &  (3.883) &  (5.947) \\[1ex]
   & CM & -~~~ & -~~~ & -~~~ & -~~~ & -~~~ & -~~~ & -~~~ & -~~~ \\
   &   &  (1.477) &  (1.207) &  (1.219) &  (0.979) &  (1.124) &  (0.909) &  (1.049) &  (1.074) \\[1ex]
   & LPS &  0.281 &  0.186 &  0.251 &  0.169 &  0.231 &  0.156 &  0.209 &  0.142 \\
   &   &  (0.290) &  (0.186) &  (0.250) &  (0.169) &  (0.227) &  (0.156) &  (0.211) &  (0.134) \\[1ex]
   & EM-I &  0.547 &  0.292 &  0.345 &  0.211 &  0.272 &  0.184 &  0.220 &  0.160 \\
   &   &  (0.318) &  (0.210) &  (0.278) &  (0.190) &  (0.237) &  (0.171) &  (0.206) &  (0.153) \\[1ex]
   & MPL-M &  0.227 &  0.144 &  0.213 &  0.134 &  0.198 &  0.123 &  0.178 &  0.111 \\
   &   &  (0.232) &  (0.146) &  (0.219) &  (0.135) &  (0.196) &  (0.121) &  (0.177) &  (0.110) \\[1ex]
   & MPL-G &  0.227 &  0.151 &  0.215 &  0.147 &  0.206 &  0.142 &  0.195 &  0.132 \\
   &   &  (0.240) &  (0.155) &  (0.231) &  (0.153) &  (0.212) &  (0.143) &  (0.200) &  (0.136) \\[1ex]
  \multicolumn{10}{l}{ }  \\
  \multicolumn{10}{l}{95\% coverage probabilities}  \\
\SizeC{\multirow{4}{*}{$h_{0}(t_{1})$}}
  & LPS &  0.916 &  0.916 &  0.938 &  0.934 &  0.939 &  0.938 &  0.939 &  0.958 \\
   & EM-I &  0.949 &  0.980 &  0.933 &  0.959 &  0.933 &  0.965 &  0.930 &  0.959 \\
   & MPL-M &  0.923 &  0.931 &  0.927 &  0.937 &  0.932 &  0.957 &  0.948 &  0.947 \\
   & MPL-G &  0.924 &  0.947 &  0.922 &  0.944 &  0.936 &  0.956 &  0.945 &  0.951 \\
      \multicolumn{10}{l}{ }  \\
\SizeC{\multirow{4}{*}{$h_{0}(t_{2})$}}
  & LPS &  0.938 &  0.940 &  0.950 &  0.958 &  0.939 &  0.962 &  0.955 &  0.960 \\
   & EM-I &  0.977 &  0.998 &  0.963 &  0.989 &  0.958 &  0.964 &  0.959 &  0.954 \\
   & MPL-M &  0.936 &  0.940 &  0.940 &  0.944 &  0.942 &  0.957 &  0.950 &  0.956 \\
   & MPL-G &  0.918 &  0.946 &  0.933 &  0.947 &  0.937 &  0.947 &  0.938 &  0.945 \\
      \multicolumn{10}{l}{ }  \\
\SizeC{\multirow{4}{*}{$h_{0}(t_{3})$}}
  & LPS &  0.942 &  0.972 &  0.954 &  0.962 &  0.941 &  0.950 &  0.954 &  0.958 \\
   & EM-I &  0.990 &  0.980 &  0.953 &  0.945 &  0.930 &  0.956 &  0.934 &  0.951 \\
   & MPL-M &  0.933 &  0.942 &  0.934 &  0.935 &  0.946 &  0.957 &  0.949 &  0.956 \\
   & MPL-G &  0.861 &  0.899 &  0.864 &  0.925 &  0.901 &  0.944 &  0.920 &  0.942 \\
      \multicolumn{10}{l}{ }  \\
  \multicolumn{10}{l}{Integrated discrepancy between $\widehat{h}_{0}(t)$ and $h_{0}(t)$ defined between 0 and the 90th percentile of $T$ }  \\
& Breslow &  7.203 &  4.331 &  3.161 &  2.752 &  2.808 &  2.662 &  2.668 &  2.594 \\
 & CM &  0.914 &  0.724 &  0.778 &  0.651 &  0.697 &  0.603 &  0.663 &  0.600 \\
 & EM-I &  0.341 &  0.204 &  0.271 &  0.171 &  0.231 &  0.146 &  0.192 &  0.125 \\
  & MPL-M &  0.192 &  0.125 &  0.182 &  0.111 &  0.161 &  0.097 &  0.145 &  0.089 \\
  & MPL-G &  0.227 &  0.143 &  0.224 &  0.144 &  0.204 &  0.130 &  0.177 &  0.112 \\
    \end{tabular}
}
\caption{Simulation 1 results for $h_{0}(t)$ for the 25th ($t_{1}$), 50th ($t_{2}$), and 75th ($t_{3}$) percentiles of T. Some results are missing for the \textit{convex minorant estimator} as no inference was not developed for this estimator, for the \textit{penalized spline estimator} as the information was missing in the article of \cite[Section 6]{CaiBet03}, and for the \textit{Breslow estimator} as we did not compute estimate the variance for this estimator.}
\label{tab-sim1h0t}
\end{table}

\begin{table} 
\centering
\SizeF{
\begin{tabular}{llrrrrrrrrr}
&    & \multicolumn{3}{c}{$\pi^{E}=0\%$} & \multicolumn{3}{c}{$\pi^{E}=25\%$}  & \multicolumn{3}{c}{$\pi^{E}=50\%$}  \\
&   & $n=100$ & $n=500$ & $n=2000$ & $n=100$ & $n=500$ & $n=2000$ & $n=100$ & $n=500$ & $n=2000$ \\
\multicolumn{11}{l}{ }  \\
\multicolumn{11}{l}{Biases}  \\
\SizeC{\multirow{5}{*}{$h_{0}(t_{1})$}}  & Breslow & -0.367 & -0.288 & -0.276 & -0.239 & -0.080 & -0.124 & -0.183 & -0.161 & -0.038 \\
   & CM & -0.019 & -0.155 & -0.371 & -0.420 & -0.464 & -0.438 & -0.290 & -0.259 & -0.284 \\
   & EM-I &  0.015 &  0.049 &  0.035 &  0.012 &  0.014 &  0.016 & -0.007 & -0.007 &  0.009 \\
   & MPL-M &  0.133 &  0.042 &  0.020 &  0.126 &  0.029 &  0.012 &  0.107 &  0.016 &  0.009 \\
   & MPL-G &  0.114 &  0.041 &  0.009 &  0.078 &  0.024 &  0.012 &  0.048 &  0.012 &  0.005 \\
     \multicolumn{11}{l}{ }  \\
\SizeC{\multirow{5}{*}{$h_{0}(t_{2})$}}  & Breslow & -0.082 & -0.177 & -0.315 & -0.048 & -0.130 & -0.186 & -0.035 & -0.069 & -0.142 \\
   & CM &  0.300 & -0.224 & -0.371 & -0.429 & -0.315 & -0.250 & -0.226 & -0.224 & -0.258 \\
   & EM-I &  0.300 &  0.036 &  0.021 &  0.201 &  0.018 &  0.012 &  0.119 &  0.009 &  0.007 \\
   & MPL-M &  0.060 &  0.032 &  0.026 &  0.051 &  0.034 &  0.016 &  0.039 &  0.025 &  0.010 \\
   & MPL-G &  0.071 &  0.064 &  0.033 &  0.071 &  0.036 &  0.011 &  0.044 &  0.018 &  0.009 \\
    \multicolumn{11}{l}{ }  \\
\SizeC{\multirow{5}{*}{$h_{0}(t_{3})$}}  & Breslow &  0.225 &  0.024 &  0.025 &  0.082 & -0.009 &  0.039 &  0.039 &  0.052 & -0.076 \\
   & CM &  0.416 & -0.411 & -0.601 & -0.424 & -0.594 & -0.618 & -0.384 & -0.504 & -0.550 \\
   & EM-I &  0.481 &  0.055 & -0.019 &  0.183 &  0.062 &  0.017 &  0.117 &  0.046 &  0.010 \\
   & MPL-M & -0.072 & -0.044 & -0.014 & -0.060 & -0.003 &  0.008 & -0.054 &  0.003 &  0.008 \\
   & MPL-G &  0.003 & -0.035 &  0.009 &  0.042 &  0.015 &  0.012 &  0.029 &  0.014 &  0.006 \\
   \multicolumn{11}{l}{ }  \\
   \multicolumn{11}{l}{Mean asymptotic (Monte Carlo) standard errors}  \\
\SizeC{\multirow{10}{*}{$h_{0}(t_{1})$}}
 & Breslow &  -~~~ & -~~~ & -~~~ & -~~~ & -~~~ & -~~~ & -~~~ & -~~~ & -~~~\\
 &  & (0.816) &  (1.873) &  (1.380) &  (1.296) &  (4.470) &  (1.913) &  (1.094) &  (1.367) &  (2.672) \\[1ex]
 & CM &  -~~~ & -~~~ & -~~~ & -~~~ & -~~~ & -~~~ & -~~~ & -~~~ & -~~~ \\
 &  & (1.437) &  (1.326) &  (0.921) &  (0.680) &  (0.551) &  (0.552) &  (1.050) &  (0.850) &  (0.683) \\[1ex]
 & EM-I &  0.789 &  0.429 &  0.193 &  0.588 &  0.255 &  0.139 &  0.495 &  0.215 &  0.124 \\
 &  &  (0.629) &  (0.366) &  (0.162) &  (0.544) &  (0.246) &  (0.132) &  (0.468) &  (0.196) &  (0.119) \\[1ex]
 & MPL-M &  0.547 &  0.224 &  0.115 &  0.478 &  0.196 &  0.101 &  0.425 &  0.176 &  0.092 \\
 &  &  (0.592) &  (0.233) &  (0.121) &  (0.488) &  (0.206) &  (0.102) &  (0.428) &  (0.178) &  (0.093) \\[1ex]
 & MPL-G &  0.567 &  0.236 &  0.123 &  0.491 &  0.213 &  0.116 &  0.437 &  0.194 &  0.104 \\
 &  &  (0.619) &  (0.253) &  (0.127) &  (0.517) &  (0.227) &  (0.116) &  (0.448) &  (0.203) &  (0.103) \\[1ex]
        \multicolumn{11}{l}{ }  \\
\SizeC{\multirow{10}{*}{$h_{0}(t_{2})$}}
 & Breslow &  -~~~ & -~~~ & -~~~ & -~~~ & -~~~ & -~~~ & -~~~ & -~~~ & -~~~\\
 &  &  (3.137) &  (3.580) &  (1.781) &  (3.157) &  (3.680) &  (2.762) &  (3.369) &  (3.747) &  (2.779) \\[1ex]
 & CM &  -~~~ & -~~~ & -~~~ & -~~~ & -~~~ & -~~~ & -~~~ & -~~~ & -~~~\\
 &  &  (4.134) &  (2.766) &  (2.980) &  (1.618) &  (1.827) &  (2.316) &  (2.361) &  (1.677) &  (2.493) \\[1ex]
 & EM-I &  2.196 &  0.988 &  0.540 &  1.4831&  0.596 &  0.320 &  1.169 &  0.486 &  0.254 \\
 &  &  (2.005) &  (0.805) &  (0.425) &  (1.504) &  (0.592) &  (0.318) &  (1.106) &  (0.439) &  (0.253) \\[1ex]
 & MPL-M &  1.061 &  0.454 &  0.231 &  0.914 &  0.397 &  0.201 &  0.812 &  0.354 &  0.180 \\
 &  &  (1.292) &  (0.482) &  (0.233) &  (0.971) &  (0.412) &  (0.199) &  (0.816) &  (0.358) &  (0.180) \\[1ex]
 & MPL-G &  1.109 &  0.481 &  0.247 &  0.982 &  0.424 &  0.222 &  0.866 &  0.379 &  0.200 \\
 &  &  (1.254) &  (0.507) &  (0.252) &  (1.057) &  (0.439) &  (0.225) &  (0.896) &  (0.380) &  (0.200) \\[1ex]
          \multicolumn{11}{l}{ }  \\
\SizeC{\multirow{3}{*}{$h_{0}(t_{3})$}}
 & Breslow &  -~~~ & -~~~ & -~~~ & -~~~ & -~~~ & -~~~ & -~~~ & -~~~ & -~~~\\
 &  & (7.646) & (13.574) &  (7.210) &  (6.311) & (12.815) &  (7.612) &  (5.466) & (11.052) &  (5.933) \\[1ex]
 & CM &  -~~~ & -~~~ & -~~~ & -~~~ & -~~~ & -~~~ & -~~~ & -~~~ & -~~~\\
 &  & (9.568) &  (3.735) &  (3.773) &  (3.341) &  (2.569) &  (2.154) &  (3.438) &  (2.215) &  (2.276) \\[1ex]
 & EM-I &  6.659 &  1.596 &  1.195 &  3.189 &  1.143 &  0.553 &  2.373 &  0.981 &  0.447 \\
 &  &  (6.979) &  (1.525) &  (0.719) &  (3.190) &  (1.119) &  (0.539) &  (2.366) &  (0.969) &  (0.434) \\[1ex]
 & MPL-M &  1.828 &  0.854 &  0.465 &  1.603 &  0.769 &  0.404 &  1.447 &  0.691 &  0.359 \\
 &  &  (2.412) &  (0.946) &  (0.482) &  (1.848) &  (0.803) &  (0.392) &  (1.586) &  (0.711) &  (0.353) \\[1ex]
 & MPL-G &  2.090 &  0.892 &  0.494 &  1.901 &  0.829 &  0.445 &  1.684 &  0.751 &  0.398 \\
 &  &  (2.379) &  (0.997) &  (0.518) &  (2.117) &  (0.868) &  (0.434) &  (1.803) &  (0.779) &  (0.390) \\[1ex]
  \multicolumn{11}{l}{ }  \\
  \multicolumn{11}{l}{95\% coverage probabilities}  \\
\SizeC{\multirow{3}{*}{$h_{0}(t_{1})$}}  & EM-I &  0.904 &  0.969 &  0.981 &  0.883 &  0.954 &  0.968 &  0.894 &  0.940 &  0.957 \\
   & MPL-M &  0.922 &  0.945 &  0.941 &  0.926 &  0.935 &  0.951 &  0.932 &  0.931 &  0.950 \\
   & MPL-G &  0.908 &  0.928 &  0.948 &  0.906 &  0.928 &  0.952 &  0.925 &  0.929 &  0.946 \\
             \multicolumn{11}{l}{ }  \\
  \SizeC{\multirow{3}{*}{$h_{0}(t_{2})$}}  & EM-I &  0.960 &  0.975 &  0.985 &  0.917 &  0.937 &  0.947 &  0.912 &  0.951 &  0.945 \\
   & MPL-M &  0.881 &  0.933 &  0.957 &  0.908 &  0.936 &  0.952 &  0.915 &  0.948 &  0.945 \\
   & MPL G &  0.884 &  0.946 &  0.948 &  0.917 &  0.939 &  0.943 &  0.916 &  0.940 &  0.947 \\
          \multicolumn{11}{l}{ }  \\
  \SizeC{\multirow{3}{*}{$h_{0}(t_{3})$}}  & EM-I &  0.970 &  0.959 &  0.998 &  0.921 &  0.953 &  0.965 &  0.899 &  0.951 &  0.954 \\
   & MPL-M &  0.735 &  0.868 &  0.925 &  0.793 &  0.919 &  0.949 &  0.821 &  0.935 &  0.953 \\
   & MPL-G &  0.829 &  0.855 &  0.930 &  0.872 &  0.924 &  0.951 &  0.875 &  0.939 &  0.949 \\
    \multicolumn{11}{l}{ }  \\
  \multicolumn{11}{l}{Integrated discrepancy between $\widehat{h}_{0}(t)$ and $h_{0}(t)$ defined between 0 and the 90th percentile of $T$}  \\
 & Breslow & 17.450 &  6.233 &  4.159 &  3.823 &  2.673 &  2.215 &  3.230 &  2.659 &  2.105 \\
 & CM &  5.173 &  2.416 &  2.564 &  2.626 &  2.457 &  2.328 &  2.558 &  2.276 &  2.072 \\
& EM-I &  5.227 &  1.150 &  0.593 &  2.274 &  0.786 &  0.392 &  1.684 &  0.655 &  0.333 \\
 & MPL-M &  1.615 &  0.727 &  0.389 &  1.352 &  0.591 &  0.292 &  1.176 &  0.518 &  0.265 \\
& MPL-G &  1.581 &  0.808 &  0.443 &  1.398 &  0.641 &  0.323 &  1.232 &  0.565 &  0.294 \\
\end{tabular}
}
\caption{Simulation 2 results for $h_{0}(t)$ for the 25th ($t_{1}$), 50th ($t_{2}$), and 75th ($t_{3}$) percentiles of T. Some results are missing for the \textit{convex minorant estimator} as no inference was not developed for this estimator, and for the \textit{Breslow estimator} as we did not estimate the variance for this estimator.}
\label{tab-sim2h0t}
\end{table}

\begin{table} 
\centering
\SizeF{
\begin{tabular}{llrrrrrrrrr}
&    & \multicolumn{3}{c}{$\pi^{E}=0\%$} & \multicolumn{3}{c}{$\pi^{E}=25\%$}  & \multicolumn{3}{c}{$\pi^{E}=50\%$}  \\
&   & $n=100$ & $n=500$ & $n=2000$ & $n=100$ & $n=500$ & $n=2000$ & $n=100$ & $n=500$ & $n=2000$ \\
\multicolumn{11}{l}{ }  \\
\multicolumn{11}{l}{Biases}  \\
\SizeC{\multirow{5}{*}{$h_{0}(t_{1})$}}
 & Breslow & -0.360 & -0.446 & -0.464 & -0.190 & -0.185 & -0.269 &  1.735 &  0.244 & -0.002 \\
 & CM &  0.194 &  0.170 &  0.114 & -0.201 & -0.167 & -0.239 &  0.086 &  0.128 &  0.059 \\
 & EM-I &  0.024 &  0.303 &  0.334 & -0.068 &  0.027 &  0.070 & -0.122 & -0.037 &  0.061 \\
 & MPL-M & -0.004 & -0.020 &  0.010 & -0.018 & -0.014 &  0.013 & -0.029 & -0.025 &  0.014 \\
 & MPL-G &  0.050 &  0.032 &  0.025 &  0.017 &  0.020 &  0.013 &  0.017 &  0.016 &  0.011 \\
       \multicolumn{11}{l}{ }  \\
\SizeC{\multirow{5}{*}{$h_{0}(t_{2})$}}
 & Breslow & -0.378 & -0.354 & -0.383 & -0.225 & -0.183 & -0.270 & -0.160 & -0.078 & -0.081 \\
 & CM &  0.591 &  0.150 &  0.120 & -0.222 & -0.111 & -0.154 & -0.012 &  0.074 & -0.004 \\
 & EM-I &  0.127 &  0.295 &  0.403 &  0.081 &  0.079 &  0.004 &  0.034 &  0.016 & -0.003 \\
 & MPL-M & -0.123 & -0.100 & -0.050 & -0.110 & -0.067 & -0.021 & -0.107 & -0.049 & -0.010 \\
 & MPL-G & -0.084 & -0.058 & -0.027 & -0.061 & -0.030 & -0.012 & -0.037 & -0.018 & -0.004 \\
      \multicolumn{11}{l}{ }  \\
\SizeC{\multirow{5}{*}{$h_{0}(t_{3})$}}
 & Breslow & -0.247 & -0.287 & -0.140 & -0.160 & -0.202 & -0.114 & -0.112 & -0.200 & -0.128 \\
 & CM &  1.024 &  0.149 &  0.011 & -0.342 & -0.597 & -0.624 & -0.399 & -0.439 & -0.423 \\
 & EM-I &  0.173 &  0.086 &  0.081 &  0.141 &  0.011 &  0.090 &  0.079 &  0.018 &  0.025 \\
 & MPL-M & -0.139 & -0.115 & -0.078 & -0.112 & -0.085 & -0.043 & -0.105 & -0.065 & -0.027 \\
 & MPL-G & -0.166 & -0.104 & -0.055 & -0.104 & -0.062 & -0.031 & -0.077 & -0.041 & -0.017 \\
   \multicolumn{11}{l}{ }  \\
   \multicolumn{11}{l}{Mean asymptotic (Monte Carlo) standard errors}  \\
\SizeC{\multirow{10}{*}{$h_{0}(t_{1})$}}
 & Breslow &  -~~~ & -~~~ & -~~~ & -~~~ & -~~~ & -~~~ & -~~~ & -~~~ & -~~~\\
 &  &  (1.746) &  (0.787) &  (0.780) &  (1.417) &  (1.182) &  (0.956) & (61.436) &  (9.629) &  (3.133) \\[1ex]
 & CM &  -~~~ & -~~~ & -~~~ & -~~~ & -~~~ & -~~~ & -~~~ & -~~~ & -~~~\\
 &  &  (1.409) &  (1.817) &  (2.620) &  (1.019) &  (0.825) &  (0.817) &  (1.478) &  (1.391) &  (1.259) \\[1ex]
 & EM-I &  0.630 &  0.415 &  0.257 &  0.432 &  0.276 &  0.129 &  0.375 &  0.195 &  0.122 \\
 &  &  (0.407) &  (0.309) &  (0.154) &  (0.388) &  (0.241) &  (0.117) &  (0.349) &  (0.182) &  (0.114) \\[1ex]
 & MPL-M &  0.397 &  0.190 &  0.110 &  0.349 &  0.170 &  0.096 &  0.314 &  0.154 &  0.087 \\
 &  &  (0.424) &  (0.201) &  (0.115) &  (0.373) &  (0.185) &  (0.098) &  (0.329) &  (0.156) &  (0.090) \\[1ex]
 & MPL-G &  0.451 &  0.211 &  0.113 &  0.398 &  0.184 &  0.098 &  0.370 &  0.170 &  0.088 \\
 &  &  (0.447) &  (0.216) &  (0.117) &  (0.418) &  (0.195) &  (0.102) &  (0.375) &  (0.170) &  (0.091) \\[1ex]
          \multicolumn{11}{l}{ }  \\
\SizeC{\multirow{10}{*}{$h_{0}(t_{2})$}}
 & Breslow &  -~~~ & -~~~ & -~~~ & -~~~ & -~~~ & -~~~ & -~~~ & -~~~ & -~~~\\
 &  &  (2.492) &  (2.276) &  (1.929) &  (2.663) &  (2.738) &  (1.948) &  (2.543) &  (2.825) &  (3.222) \\[1ex]
 & CM &  -~~~ & -~~~ & -~~~ & -~~~ & -~~~ & -~~~ & -~~~ & -~~~ & -~~~\\
 &  &  (5.265) &  (4.801) &  (3.890) &  (1.664) &  (1.709) &  (1.655) &  (1.856) &  (2.125) &  (1.704) \\[1ex]
 & EM-I &  0.970 &  0.619 &  0.515 &  0.818 &  0.493 &  0.231 &  0.751 &  0.377 &  0.192 \\
 &  &  (0.981) &  (0.460) &  (0.241) &  (0.811) &  (0.395) &  (0.213) &  (0.714) &  (0.318) &  (0.188) \\[1ex]
 & MPL-M &  0.632 &  0.293 &  0.169 &  0.558 &  0.268 &  0.151 &  0.500 &  0.249 &  0.138 \\
 &  &  (0.686) &  (0.319) &  (0.180) &  (0.603) &  (0.301) &  (0.156) &  (0.538) &  (0.263) &  (0.146) \\[1ex]
 & MPL-G &  0.676 &  0.316 &  0.173 &  0.616 &  0.286 &  0.154 &  0.577 &  0.266 &  0.141 \\
 &  &  (0.708) &  (0.329) &  (0.178) &  (0.661) &  (0.306) &  (0.157) &  (0.603) &  (0.267) &  (0.147) \\[1ex]
            \multicolumn{11}{l}{ }  \\
\SizeC{\multirow{10}{*}{$h_{0}(t_{3})$}}
 & Breslow &  -~~~ & -~~~ & -~~~ & -~~~ & -~~~ & -~~~ & -~~~ & -~~~ & -~~~\\
 &  &  (2.764) &  (2.655) &  (7.332) &  (4.076) &  (3.246) &  (6.000) &  (4.199) &  (3.127) &  (5.241) \\[1ex]
 & CM &  -~~~ & -~~~ & -~~~ & -~~~ & -~~~ & -~~~ & -~~~ & -~~~ & -~~~\\
 &  & (10.753) &  (5.271) &  (5.261) &  (2.779) &  (1.310) &  (0.966) &  (1.755) &  (1.450) &  (1.769) \\[1ex]
 & EM-I &  1.577 &  0.814 &  0.463 &  1.282 &  0.584 &  0.351 &  1.134 &  0.497 &  0.296 \\
 &  & (1.656) &  (0.571) &  (0.256) &  (1.318) &  (0.477) &  (0.312) &  (1.012) &  (0.452) &  (0.274) \\[1ex]
 & MPL-M &  0.938 &  0.422 &  0.227 &  0.832 &  0.376 &  0.205 &  0.741 &  0.345 &  0.188 \\
 &  & (1.064) &  (0.459) &  (0.238) &  (0.949) &  (0.420) &  (0.216) &  (0.839) &  (0.373) &  (0.201) \\[1ex]
 & MPL-G &  0.912 &  0.429 &  0.233 &  0.856 &  0.390 &  0.210 &  0.790 &  0.362 &  0.192 \\
 &  &  (0.999) &  (0.455) &  (0.235) &  (0.968) &  (0.416) &  (0.215) &  (0.847) &  (0.380) &  (0.201) \\[1ex]
            \multicolumn{11}{l}{ }  \\
  \multicolumn{11}{l}{Empirical 95\% coverage probabilities}  \\
\SizeC{\multirow{3}{*}{$h_{0}(t_{1})$}}
 & EM-I &  0.947 &  0.925 &  0.765 &  0.874 &  0.940 &  0.953 &  0.839 &  0.911 &  0.949 \\
   & MPL-M &  0.859 &  0.917 &  0.942 &  0.881 &  0.909 &  0.947 &  0.879 &  0.934 &  0.938 \\
   & MPL-G &  0.899 &  0.947 &  0.940 &  0.903 &  0.935 &  0.938 &  0.905 &  0.950 &  0.942 \\
               \multicolumn{11}{l}{ }  \\
\SizeC{\multirow{3}{*}{$h_{0}(t_{2})$}}
& EM-I &  0.910 &  0.821 &  0.691 &  0.881 &  0.972 &  0.962 &  0.895 &  0.961 &  0.953 \\
   & MPL-M &  0.786 &  0.810 &  0.859 &  0.802 &  0.849 &  0.920 &  0.807 &  0.890 &  0.934 \\
   & MPL-G &  0.828 &  0.896 &  0.922 &  0.851 &  0.898 &  0.928 &  0.871 &  0.939 &  0.938 \\
            \multicolumn{11}{l}{ }  \\
\SizeC{\multirow{3}{*}{$h_{0}(t_{3})$}}
& EM-I &  0.948 &  0.949 &  0.975 &  0.921 &  0.953 &  0.934 &  0.935 &  0.944 &  0.968 \\
   & MPL-M &  0.757 &  0.775 &  0.772 &  0.793 &  0.801 &  0.877 &  0.784&  0.845 &  0.893 \\
   & MPL-G &  0.731 &  0.801 &  0.850 &  0.805 &  0.863 &  0.910 &  0.823 &  0.882 &  0.914 \\
      \multicolumn{11}{l}{ }  \\
  \multicolumn{11}{l}{Integrated discrepancy between $\widehat{h}_{0}(t)$ and $h_{0}(t)$ defined between 0 and the 90th percentile of $T$}  \\
& Breslow &  0.810 &  0.739 &  0.573 &  0.789 &  0.704 &  0.521 &  0.775 &  0.688 &  0.495 \\
& CM &  1.277 &  0.691 &  0.589 &  0.705 &  0.588 &  0.573 &  0.623 &  0.529 &  0.522 \\
& EM-I &  0.383 &  0.223 &  0.207 &  0.327 &  0.153 &  0.094 &  0.288 &  0.136 &  0.077 \\
 & MPL-M &  0.326 &  0.161 &  0.089 &  0.287 &  0.139 &  0.070 &  0.259 &  0.121 &  0.063 \\
 & MPL-G &  0.317 &  0.153 &  0.079 &  0.285 &  0.130 &  0.068 &  0.255 &  0.116 &  0.060 \\
\end{tabular}
}
\caption{Simulation 3 results for $h_{0}(t)$ for the 25th ($t_{1}$), 50th ($t_{2}$), and 75th ($t_{3}$) percentiles of T. Some results are missing for the \textit{convex minorant estimator} as no inference was not developed for this estimator, and for the \textit{Breslow estimator} as we did not estimate the variance for this estimator.}
\label{tab-sim3h0t}
\end{table}

\end{document}